\documentclass[prb,showpacs,twocolumn,superscriptaddress]{revtex4-1}
\usepackage{amsmath,latexsym,bm,psfrag,color,psfrag}
\usepackage[pdftex]{graphicx}

\begin{document}

\title{Model of two-dimensional electron gas
formation at ferroelectric interfaces}

\author{P. Aguado-Puente}\email{p.aguado@nanogune.eu}
\affiliation{Donostia International Physics Center, 
         Paseo Manuel de Lardizabal 4, 20018 San Sebasti\'an, Spain}
\affiliation{CIC Nanogune, Tolosa Hiribidea 76, 20018 San Sebasti\'an, Spain}

\author{ N. C. Bristowe}
\affiliation{Theoretical Materials Physics, University of Li\`ege,
         B-4000 Sart-Tilman, Belgium}
\affiliation{Department of Materials, Imperial College London, 
         London SW7 2AZ, UK}
         
 \author{ B. Yin}
\affiliation{Department of Engineering Mechanics, Zhejiang University,
         Hangzhou 310027, China} 
\affiliation{CIC Nanogune, Tolosa Hiribidea 76, 20018 San Sebasti\'an, Spain}
         
\author{R. Shirasawa}
\affiliation{Department of Earth Sciences, University of Cambridge, 
         Downing Street, Cambridge CB2 3EQ, UK}
\altaddress[Current address: ]{Sony Corporation, Atsugi-shi,
                Kanagawa, 243-0021, Japan}
\author{Philippe Ghosez}
\affiliation{Theoretical Materials Physics, University of Li\`ege,
             B-4000 Sart-Tilman, Belgium}
\author{P. B. Littlewood}
\affiliation{Physical Sciences and Engineering,
             Argonne National Laboratory, 
             Argonne, Illinois 60439, USA}   
\affiliation{University of Chicago,
             James Frank Institute, 
             Chicago, Illinois 60637, USA}
\author{Emilio Artacho}
\affiliation{Donostia International Physics Center, 
         Paseo Manuel de Lardizabal 4, 20018 San Sebasti\'an, Spain}
\affiliation{CIC Nanogune, Tolosa Hiribidea 76, 20018 San Sebasti\'an, Spain}
\affiliation{Theory of Condensed Matter,
             Cavendish Laboratory, University of Cambridge, 
             J. J. Thomson Ave, Cambridge CB3 0HE, UK}
\affiliation{Basque Foundation for Science Ikerbasque, 48013 Bilbao, Spain}

\date{\today}

\begin{abstract}
The formation of a two-dimensional electron gas at oxide interfaces
as a consequence of polar discontinuities
has generated an enormous amount of activity due to the variety of
interesting effects it gives rise to. Here we study under what circumstances
similar processes can also take place underneath ferroelectric thin films. 
We use a simple Landau model to demonstrate that 
in the absence of extrinsic screening mechanisms
a monodomain phase can be stabilized in ferroelectric films
by means of an electronic reconstruction.
Unlike in the LaAlO$_3$/SrTiO$_3$ heterostructure, the emergence 
with thickness of the 
free charge at the interface is discontinuous.
This prediction is confirmed by performing first principles simulations
of free standing slabs of PbTiO$_3$.
The model is also used to predict the response of the system to an 
applied electric field, demonstrating that the two-dimensional 
electron gas can be switched on and off discontinuously
and in a non-volatile fashion.
Furthermore, the reversal of the polarization can be used to switch
between a two-dimensional electron gas and a two-dimensional hole gas, 
which should, in principle, have very different transport properties.
We discuss the possible formation of polarization domains and how
such configuration competes with the  spontaneous accumulation of
free charge at the interfaces.

\end{abstract}

\maketitle

\section{Introduction}

After the discovery of the formation of a two-dimensional electron gas (2DEG) 
at some oxide interfaces \cite{Ohtomo-02,Ohtomo-04} it was immediately 
realized that this system possessed a number of potential applications. 
The great efforts devoted to the investigation 
of this unexpected phenomenon have indeed
yielded a fantastic variety of functionalities that can be tailored
in these systems such as superconductivity \cite{Reyren2007} or enhanced
capacitance. \cite{Li2011}
Furthermore, the occurrence of a 2DEG in a perovskite system
opens possibilities for coupling such 2DEG to other
interesting properties commonly found in different perovskites,
from high-$T_c$ superconductivity to multiferroicity.

The driving force for the formation of the 2DEG at oxide interfaces,
such as the LaAlO$_3$/SrTiO$_3$ heterostructure,
is the polar discontinuity at the boundary 
between two materials with different formal
polarizations. \cite{Stengel2009a} This polarization mismatch
has a huge electrostatic cost and can favor the formation of 
free charge that accumulates at the interface in order to 
screen the discontinuity, the process sometimes referred to
as the ``polar catastrophe''. \cite{Nakagawa2006}
One interesting aspect of this phenomenon is that the 
magnitude of the polar discontinuity can be tuned in a
number of different ways: using different interface 
orientations, \cite{Annadi2013} alloying the polar material to effectively 
change its polarization \cite{Reinle-Schmitt2012} or playing with the electrostatic 
boundary conditions of the system. \cite{Thiel2006}
One possible way to manipulate the electrostatic boundary conditions
is through the coupling with ferroelectricity. 
This strategy has been 
considered in the past, 
\cite{Bristowe2009, Bark2011, Tra2013, Niranjan2009, Wang2009} 
since the spontaneous polarization
of the ferroelectric material could be used to tune the 
polar mismatch at the interface.
Indeed, the manipulation of the 2DEG in
LaAlO$_3$/SrTiO$_3$ using ferroelectricity
has already been achieved experimentally in different ways. In Ref. 
\onlinecite{Bark2011} epitaxial strain was used to 
induce a ferroelectric phase transition in  
SrTiO$_3$, whose spontaneous polarization 
was observed to partially screen the polar discontinuity, thus 
reducing the carrier concentration and increasing the critical 
thickness of LaAlO$_3$ for the formation of the 2DEG. 
Alternatively, in Ref. \onlinecite{Tra2013} V. T. Tra and coauthors used a 
ferroelectric over-layer to top-gate the LaAlO$_3$/SrTiO$_3$
heterostructure, being able to induce a metal-insulator 
transition at the interface in a non-volatile way by 
switching the polarization of the ferroelectric.
A more radical approach is directly to substitute the polar LaAlO$_3$
by a ferroelectric material and use the spontaneous polarization of the 
ferroelectric as the source for the polar discontinuity.
This possibility has already been explored from
first principles. \cite{Niranjan2009, Wang2009, Fredrickson2015} 
In Refs. \onlinecite{Niranjan2009, Wang2009} 
it was shown that the 2DEG could be manipulated with the 
ferroelectric polarization in symmetric KNbO$_3$/ATiO$_3$ 
(A$=$Sr, Ba, Pb) heterostructures. 
However, the non-stoichiometry of the simulated geometry implied that the
interfaces studied in those works were metallic \emph{by construction}. In fact it was 
later demonstrated \cite{Garcia-Fernandez2013} that since centrosymmetric 
KNbO$_3$ is polar with a formal polarization of 
half a quantum of polarization (modulo a quantum of polarization), 
just like LaAlO$_3$, when a $[001]$ interface between this material 
and a non-polar one is grown the \emph{ferroelectric} polarization of KNbO$_3$
tends to compensate the polarity of the interface. 
As a result, the polarization of the KNbO$_3$ layer
is pinned and its formal value in the ferroelectric 
ground state is approximately zero (up to a quantum of
polarization) rendering any screening mechanism unnecessary.

Instead, a ferroelectric with a non-polar 
centrosymmetric high-temperature phase
should be used. 
In that case the polarization in the 
ferroelectric phase is not intrinsically compensated.
For a free-standing slab or a thin film of such material on top of
an insulating, non-polar substrate (such as the common SrTiO$_3$), the switchable
polarization of the ferroelectric could be used to manipulate the electrostatic 
boundary conditions at the interface and possibly induce the formation of a 2DEG.
This, of course, would only be possible if such configuration, a monodomain
ferroelectric phase screened by a 2DEG, is stable,
because, unlike LaAlO$_3$, for a ferroelectric thin 
film the system has alternative routes available to minimize or avoid the
polarization mismatch. For one, in the absence of a screening mechanism other 
than the accumulation of free charge at the interface 
depolarization effects might render the paraelectric configuration as the 
only stable homogeneous phase of the system.
But most notably the system
can break into polarization domains.
Strikingly, reports of monodomain phases in ferroelectric thin films
on insulating substrates are not rare in the experimental literature,
\cite{Thompson1997, Bedzyk2000, Streiffer2002, Fong2004a, 
Fong2005, Tenne2009, Dubourdieu2013a} even though a simple
electrostatic analysis reveals that such configuration can only be stable
if free charge accumulates at the interfaces. In this geometry, interface or
surface atomic reconstructions, or simple adsorption
of ionic species to the surface do not provide the necessary 
screening. \cite{Bristowe2014} A transfer of charge from 
the surface to the interface or vice versa is needed, 
which, as in the case 
of LaAlO$_3$/SrTiO$_3$, might come from different sources, such as 
an electronic reconstruction or redox processes.
In fact, a very recent first-principles study
has shown that electronic reconstruction can stabilize a
polarization in a BaTiO$_3$ thin film on top of SrTiO$_3$. \cite{Fredrickson2015}
In that work, simulations performed for a specific 
thickness of the ferroelectric film showed that, even if the ground
state of the system was the paraelectric phase, a configuration with a
finite polarization pointing towards the substrate and a 2DEG at the interface
was metastable.
Neither the polarization reversal nor a metal-insulator transition with thickness
or electric field
could be demonstrated, but the work of Ref. \onlinecite{Fredrickson2015}
together with all the previous arguments suggest that the formation
and manipulation of a 2DEG at ferroelectric interfaces might indeed be possible.

The complex phenomenology that is expected for these systems cannot be 
explored exclusively within a first principles approach.
The relative stability of the polar configuration with
respect to competing phases, the evolution of such competition with the thickness, or   
the response of the system to an external electric field 
(which is the main interest of having a 2DEG in ferroelectric films) are
issues of prime importance for which a systematic 
first principles analysis is today unfeasible.
In this work we use a phenomenological model, supported 
by first principles calculations, to confirm that
ferroelectricity can be used to induce the formation 
of two-dimensional electron and holes gases at the interface with
non-polar substrates. We discuss the conditions for the stability of 
such configuration,  its coupling with external electric fields --
which gives rise to a discontinuous switching (on and off) of the gas--,
and its competition with alternative screening mechanisms
such as the formation of polydomain phases.
The paper is organized as follows: In Sec. \ref{sec:Formation} we present the
model and use it to predict the range of stability of a 2DEG in a prototypical
system, in Sec. \ref{sec:efield} we analyze the interaction 
with an external electric field and in Sec. \ref{sec:Discussion} we discuss
the implications of the results, in particular how they are affected
by the competition with polydomain phases and what is
the expected phenomenology in the case of the recently proposed
hyperferroelectric materials. \cite{Garrity2014}

\section{Formation of a 2DEG at ferroelectric interfaces}
\label{sec:Formation}
\subsection{A simple model}
\label{sec:theModel}

Phenomenological models have been successfully used to 
rationalize the formation of a 2DEG at polar interfaces
between paraelectric materials. \cite{Bristowe2011,Bristowe2014} 
These models allow to assess the viability of 
different processes able to screen the polar discontinuity by injecting free 
charge into the interfaces/surfaces.
Here we use the same formalism to explore the formation 
of a 2DEG at ferroelectric thin films.
We first consider the possible transition from paraelectric to
ferroelectric with a 2DEG, neglecting the competition with 
the formation of domains (discussed in Sec. \ref{sec:poly}).

We start assuming that no extrinsic mechanisms contribute
to the screening, thus the electronic reconstruction 
(for which electrons from one surface/interface 
are transferred to the opposite one to screen the polarization of the 
film) is the only possible source of free charge. \cite{Bristowe2014}
Throughout this paper, for the sake of conciseness,
we only speak about 2DEG, but it should be noted that 
under the assumption of electronic reconstruction the formation
of a 2DEG implies the appearance of a corresponding two-dimensional
hole gas (2DHG) at the opposite
interface or surface.
Later on we discuss how this model 
can also be used to describe the basic behavior of the system 
when the free carriers are provided by surface redox processes
(like formation of charged defects or adsorption of chemical species). 

\begin{figure}[]
  \begin{center}
    \includegraphics[clip,width=\columnwidth]{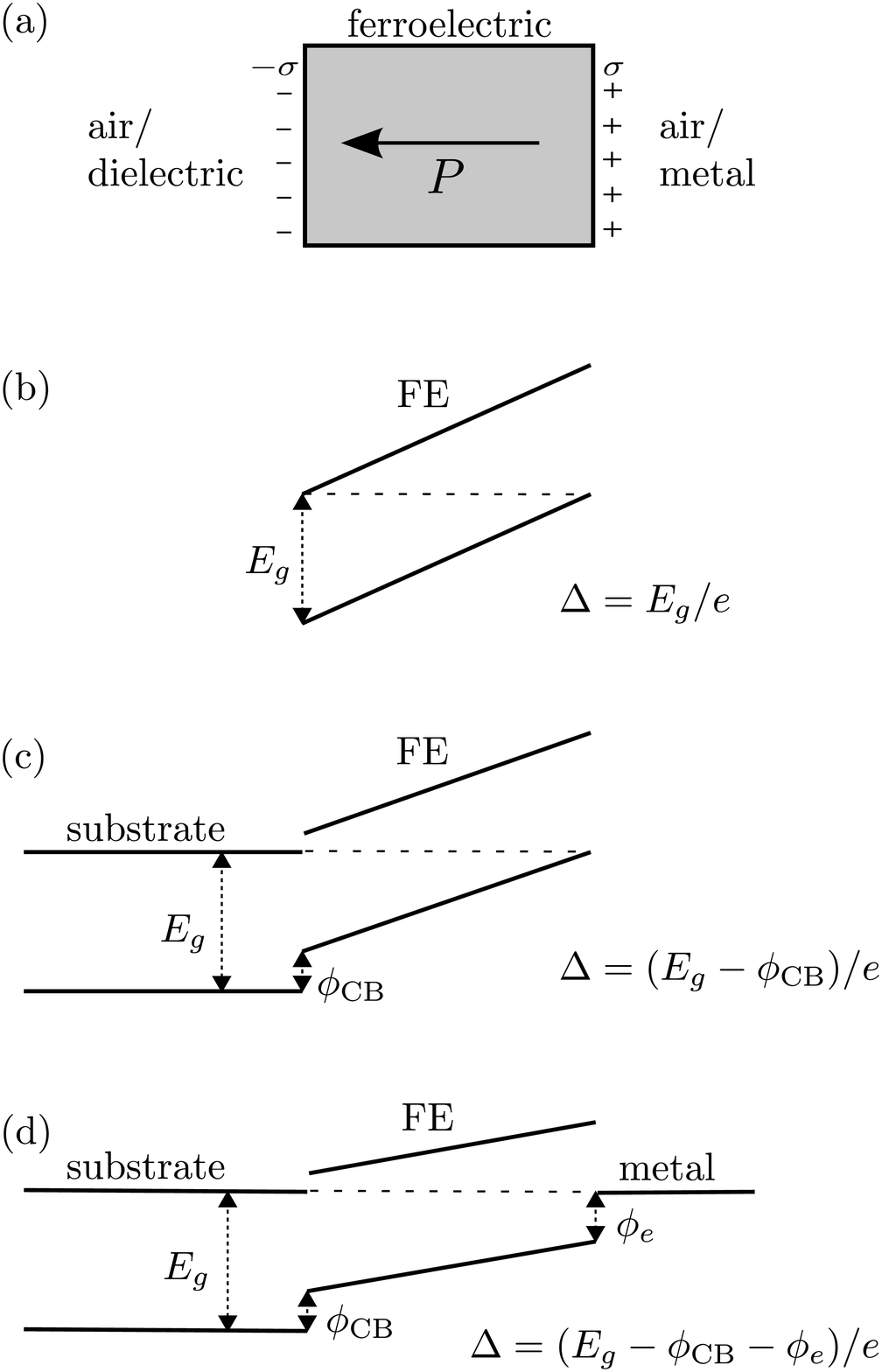}
  \end{center}
  \caption{(a) Schematic diagram of the geometry of the system. 
  The sign criterion in all the equations throughout the paper
  assume absolute values of both the 
  polarization and the free charge density, and their relative
  orientation is the one given in this figure. (b) to (d), schematic
  band alignment for a ferroelectric thin film in various configurations.
  The corresponding value of the relevant gap, $\Delta$; 
  the band offset at the interface, $\phi_{CB}$; and the Shottky
  barrier for electron in the presence of an electrode, $\phi_e$, are indicated in
  each case.
  }
  \label{fig:geom}
\end{figure}

If the 2DEG forms as a result of electronic reconstruction, 
the free energy \emph{per unit volume} of a ferroelectric thin film of thickness
$d$ reads
\begin{equation}
G = U + {\sigma \Delta \over d} + {\sigma^2 \over 2gd} + 
    {1 \over 2 \varepsilon_0} (\sigma -P )^2 \; .
 \label{eq:Ginitial}
\end{equation}
In this expression $U$ is the free energy of the bulk ferroelectric 
at zero field, that depends
on the polarization $P$.
The next two terms on the right-hand side of Eq. \ref{eq:Ginitial}
account for the energy cost of promoting electrons from the top
of the valence band to the bottom of the conduction band.
The second term of the equation corresponds to the cost of the
charge transfer across the gap, where 
$\sigma$ is the surface density of free charge,
and $\Delta$ is the ``relevant band gap'' of the system in units of a voltage.
Neglecting surface effects, $\Delta$ is equal to the band gap 
of the ferroelectric in the case of a 
free standing slab [see Fig. \ref{fig:geom}(b)], but in general
its value depends on the configuration of the heterostructure. 
Fig. \ref{fig:geom} illustrates different cases, where $\Delta$ is 
calculated from the band gap of the constituent materials, the
band alignment across interfaces and the Fermi level of a top electrode
(if present). If band bending or any other modification
of the electronic structure occurs at the interfaces, it 
should also be taken into account. Throughout this paper, for
the numerical estimations we will consider the simplest approximation
and use the band gap of the bulk ferroelectric for $\Delta$.
The third term in Eq. \ref{eq:Ginitial}
takes into account the energy associated to the filling of the bands
(the conduction band with electrons and the valence band with holes). 
This energy cost is associated with a finite density of states. 
The ``reduced density of states'' \cite{Bristowe2014} used in 
Eq. \ref{eq:Ginitial} is calculated as $g = (g_e g_h) / (g_e+g_h)$
and is expressed in units of charge
squared per units of area and energy, with $g_e$ and $g_h$ 
being the density of states (DOS) for electrons and holes respectively
(we take the densities of states as constants, as in a free electron
gas in 2D). 
The last term in Eq. \ref{eq:Ginitial} 
corresponds to the electrostatic energy of the remnant depolarizing field, and 
constitutes the driving force for the formation of the 2DEG.

There are several subtleties regarding Eq. \ref{eq:Ginitial} that should 
be noted.
Firstly, our Landau-type model is restricted to an out-of-plane 
polarization; it does not include the possibility of in-plane 
polarization nor any explicit strain dependence.
Some simple mechanical boundary conditions, such as
epitaxial strain, can be implicitly accounted for
by a renormalization of the coefficients in the 
expansion of $U$.\cite{Pertsev1998a}
The generalization needed to explicitly include these additional degrees 
of freedom is nevertheless trivial: $U$ would depend on the three 
components of the polarization and on the strain,  
the out-of-plane component of $P$ being the only relevant one 
in the depolarization term. 
Such a model would allow for a rotation of the polarization in-plane, 
which in some cases might be a competing mechanism to avoid the 
electrostatic cost associated to a discontinuity of the 
out-of-plane polarization. 
Such possibility is however not discussed here because the 
in-plane epitaxial compressive strain imposed by the substrate 
hinders the stability of an in-plane polarization in the 
prototypical systems of interest (such as epitaxial 
PbTiO$_3$ or BaTiO$_3$ on SrTiO$_3$).
Secondly, in Eq. \ref{eq:Ginitial} we assume that the distance between 
electron and hole layers is large enough that 
exchange interactions, as well as excitonic binding, can be 
disregarded.
Finally, throughout this paper
both $\sigma$ and $P$ are the magnitudes of the physical quantities and the
signs of the different terms in all the equations
are valid for the geometry and relative orientation of $\sigma$ and $P$
depicted schematically in Fig. \ref{fig:geom}(a). 
For an arbitrary sign of the free charge  
with respect to the polarization one should take into account that 
the term corresponding to the band gap energy should read 
$|\sigma|\Delta$. Furthermore, since the orientation
of the polarization has a strong influence on the
band alignment of an interface, the relevant gap 
may, in general, have a different value for opposite 
orientations of the polarization. 
This difference, together with the choice of a suitable thickness, 
might be exploited to switch on and off the 2DEG,
as will be discussed below. 

For prototypical ferroelectric materials \emph{at zero 
electric field} the relevant free energy can be expressed as a Landau 
expansion in terms of a single order parameter responsible for the 
ferroelectric phase transition $\eta$, which for these materials
consists of a soft mode associated to a collective shift
of the oxygen cage with respect to the cations. Alternatively
one can use the polarization associated with this mode,
$P_\eta={1 \over \Omega} Z_\eta^* \eta$, where $\Omega$ is the unit
cell volume of the ferroelectric and $Z_\eta^*$ is the Born effective
charge associated with the mode $\eta$. 
The Landau expansion of the free energy per unit volume
in terms of this polarization is 
expressed as, 
\begin{equation}
U_0 = {1\over 2} a (T-T_\mathrm{C}) P_\eta^2 + {1\over 4} b P_\eta^4 \ 
+ \mathcal{O}(P_\eta^6),
\label{eq:Landau1}
\end{equation}
where $T_\mathrm{C}$ is the Curie temperature of the material. 
For materials with a second-order phase transition with the temperature, 
the coefficients of $P_\eta^4$ and larger-order terms are positive and
the energy expansion
of Eq. \ref{eq:Landau1} may be truncated at the quartic term. 
This allows most of the analysis that follows to be 
done in terms of analytical expressions and provides a
direct relationship between the constants in the Landau expansion and
common physical properties such as the spontaneous polarization
and susceptibility.
A more general discussion should take into account higher order 
terms and, as in the case of improper ferroelectrics, the coupling 
of polar modes with non polar distortions.
However here we restrict ourselves to materials which can be described
by the expression in Eq. (2), since this includes some
prototypical systems such as BaTiO$_3$ and PbTiO$_3$ under
compressive epitaxial strain \cite{Pertsev1998a} 
(again, this is the case if either of
these materials is grown on a SrTiO$_3$ substrate).
In fact, since the phenomenology described in this work is a 
consequence of the ``double-well'' shape of the free energy 
as a function of the polarization of
ferroelectrics, the behavior
derived from the model should also be qualitatively 
valid for ferroelectrics
with a first-order phase transition provided that $T\ll T_\mathrm{C}$.

At a given temperature below the phase transition (and leaving
temperature aside for the time being), we rewrite Eq. \ref{eq:Landau1} as 
\begin{equation}
U_0 = {1 \over 2 \varepsilon_0 \chi_\eta} \left ( {1\over 4} {P_\eta^4 \over P_S^2} 
- {1\over 2} P_\eta^2 \right ) \; , 
\label{eq:U_0}
\end{equation}
where  $P_S = [(a/b) (T_\mathrm{C}-T)]^{1/2}$ is the spontaneous polarization
in the absence of a depolarizing field, and 
$\varepsilon_0 \chi_\eta =  [ 2 a (T_\mathrm{C}-T)]^{-1}$ 
is the contribution of $\eta$ to the polarizability 
around $P_S$. $\chi_\eta$ would correspond to the curvature 
around the minimum of the
double well energy curve $U_0(P_\eta)$, 
typically obtained from first principles
performing a series of frozen phonon calculations at zero field.
The expression in Eq. \ref{eq:U_0}, however, is only valid 
at zero field, since it does not include the extra 
polarization of the electrons and other phonons in arbitrary electrostatic
boundary conditions. In general, the Landau expansion
of the energy would be 
\begin{equation}
U = {1 \over 2 \varepsilon_0 \chi_\eta} \left ( {1\over 4} {P_\eta^4 \over P_S^2} 
- {1\over 2} P_\eta^2 \right ) + {1 \over 2\varepsilon_0\chi_\infty}P_e^2 \; , 
\label{eq:U}
\end{equation}
where the total polarization of the material is
\begin{equation}
P = P_\eta + \varepsilon_0\chi_\infty\mathcal{E} = P\eta + P_e. 
\label{eq:U}
\end{equation}
Note that $P_\eta$ already includes a contribution from the electronic 
cloud contained in $Z_\eta^*$, since this is a dynamical charge
that takes into account the deformation of the
electronic charge density with the amplitude of the polar distortion
at zero field. 
Accordingly $P_e$ and $\chi_\infty$ are the extra polarization
and susceptibility due to the presence of a finite electric field.
$P_e$ and $\chi_\infty$ account mainly for the polarizability
of the electronic cloud, thus we will refer to them as \emph{electronic}
polarization and susceptibility throughout the paper; however, strictly
speaking, these two terms also include the contribution 
of hard modes of the lattice. \cite{Tagantsev1986a,Stengel2012}
Using the electrostatic boundary conditions of our problem we can express $P_e$
in terms of the total polarization, $P$, as
\begin{equation}
 P_e = \varepsilon_0\chi_\infty\mathcal{E} = 
      \varepsilon_0\chi_\infty {\sigma - P \over \varepsilon_0} =
      \chi_\infty\left(\sigma - P \right),
 \label{eq:Pe}
\end{equation}
which, in turn, can be written as a function of the zero-field
polarization as
\begin{equation}
 P = {P_\eta + \chi_\infty \sigma \over \varepsilon_\infty},
 \label{eq:P}
\end{equation}
where $\varepsilon_\infty = \chi_\infty +1$ is the electronic (or background) contribution
to the relative permittivity of the ferroelectric.
Using Eq. \ref{eq:U} through \ref{eq:P}, Eq. \ref{eq:Ginitial} 
transforms into
\begin{equation}
\begin{aligned}
 G = & {1 \over 2\varepsilon_0 \chi_\eta}
     \left({1\over 4} {P_\eta^4 \over P_S^2} - {1 \over 2}P_\eta^2\right) \\
     & + {1 \over 2\varepsilon_0 \varepsilon_\infty} \left(\sigma - P_\eta\right)^2 
     + {\Delta\sigma\over d} + {\sigma^2 \over 2gd}.
 \label{eq:Gfinal}
\end{aligned}
\end{equation}
Note that the second term in Eq. \ref{eq:Gfinal} looks very similar
to the last term in Eq. \ref{eq:Ginitial} and consequently it could
be misunderstood as the energy due to the depolarizing field, but in fact
it contains that contribution as well as the one due to the 
electronic polarization $P_e$. 

Equation \ref{eq:Gfinal} can be used to find the equilibrium polarization
and surface/interface free charge in a ferroelectric thin film. We will
neglect for the moment the influence of the DOS assuming 
that $g$ is relatively large. Since the DOS term is inversely proportional 
to $g$, it decays rapidly for relatively large, but realistic, values of the DOS.
Therefore, the approximation $g \rightarrow \infty$ can be made without 
significantly affecting the qualitative behavior of the system 
and simplifying the analysis that follows.
%
%
Under this approximation, 
the two equilibrium conditions $\partial G / \partial P_\eta =
\partial G / \partial \sigma =0 $ yield
the system of equations
\begin{eqnarray}
 \label{eq:dG/dP}
 {P_\eta^3 \over P_S^2}  - P_\eta - {2 \chi_\eta \over \varepsilon_\infty}
 (\sigma - P_\eta)  &= 0 \\
 \label{eq:dG/dsigma}
 {\Delta\over d} + \frac{1}{\varepsilon_0 \varepsilon_\infty}(\sigma - P_\eta) &= 0 \, .
\end{eqnarray}
under the constraint of $\sigma \ge 0$.
In the limit of large film thickness the solutions of these
two equations are $\sigma = P_\eta$ and $P_\eta =\{-P_S, 0, P_S \}$; and,
using Eq. \ref{eq:P}, $P = P_\eta$.
This just means that in thick ferroelectric films, the bulk tendency
dominates, polarizing accordingly, and the free charge at the surfaces
just follows, precisely canceling the depolarizing field
(as in shorted boundary conditions).
As $d$ decreases from $\infty$, $P_\eta$ diminishes, but 
$\sigma$ diminishes faster, leaving part of the polarization
uncompensated (the energy cost of the depolarizing field
does not completely overwhelm the energy cost of 
transferring charge from the valence to the conduction band, $\Delta$). 

In order to study the thickness dependence of the equilibrium
polarization and free charge density, 
Eq. \ref{eq:dG/dP} and \ref{eq:dG/dsigma} can be combined to
obtain the condition
\begin{equation}
{P_\eta^3 \over P_S^3}  - {P_\eta \over P_S}  + {l \over d}  = 0, 
\label{cubic2}
\end{equation}
where
\begin{equation}\label{lferro}
l = 2 \chi_\eta {\varepsilon_0 \Delta \over P_S}, 
\end{equation}
is a characteristic length scale for a given set of parameters.
In the plot of $P$ or $\sigma$ versus $d$, the bulk spontaneous
polarization $P_S$ defines the scale for $P$ and $\sigma$, and
the length $l$ defines the scale for film thickness, $d$.

\begin{figure}
\begin{center}
\includegraphics[clip,width=\columnwidth]{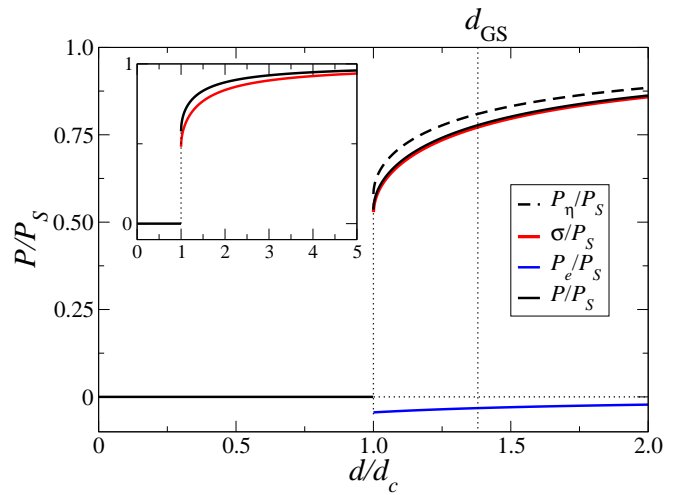}
\caption{\label{Ferro}  Total polarization (black solid line), 
zero-field contribution (dashed line), electronic
contribution (solid blue line) and surface free
charge (solid red line)
versus ferroelectric film thickness ($d$). 
  $P_S$ is the spontaneous polarization of the
bulk ferroelectric material, $d_c$ is the critical thickness
for the onset of ferroelectric metastability, and $d_\mathrm{GS}$ is the
thickness at which the ferroelectric configuration becomes the
ground state of the system.
Inset: Same plot with $\chi_\eta=2$ and $\varepsilon_\infty = 1$ to
highlight the convergence of $\sigma$ towards $P$ (in this situation
$P=P_\eta$).
}
\end{center}
\end{figure}

  The evolution of both $P$ and $\sigma$ is shown in 
Fig.~\ref{Ferro} as a function of $d$.
For this plot we used the parameters for PbTiO$_3$ obtained
from first principles
\footnote{We are using parameters corresponding to PbTiO$_3$
strained in-plane to an implicit SrTiO$_3$ substrate. The corresponding
parameters, obtained from first principles calculations are: 
$P_S = 0.78$ C/m$^2$, 
$\chi_\eta = 26$, 
$\varepsilon_\infty = 7$,
$g_e/e^2 = 1.2\cdot 10^{37}$ m$^{-2}$J$^{-1}$, and  
$g_h/e^2 = 2.5\cdot 10^{37}$ m$^{-2}$J$^{-1}$.
For the study of the competition with a polydomain phase we use 
$\Sigma = 130$ mJ/m$^2$ (from Ref. \onlinecite{Meyer2002}),
$\varepsilon_z = \varepsilon_\infty + (\chi_\eta+1)=35$, and
$\varepsilon_x = 185$.},
except for $\Delta$, for which we used the experimental band gap,
i.e. $e\Delta=3.6$ eV.
  The thick film limit (large $d$) displays what was described
before, namely, ($i$) $P$ tends to the bulk value $P_S$, and 
($ii$) the free charge density $\sigma$ tends to
screen the polarization.
The polarization then diminishes for thinner films
until a critical thickness 
\begin{equation}\label{dcferro}
d_c = l {3 \sqrt{3} \over 2},
\end{equation}
where a discontinuous jump in all magnitudes occur,
and below which $P=\sigma=0$.
  The value of the polarization at the critical thickness
is 
\begin{equation}
\label{eq:Pc}
P_\eta^c = {P_S \over \sqrt{3} } \, ,
\end{equation}
independent of other parameters.
The implications of Eq. \ref{dcferro} and \ref{eq:Pc}
are quite remarkable, in the sense that these expressions suggest
that this is a rather general behavior for ferroelectric materials [at least for 
those which respond to the energy expression of Eq. \ref{eq:Landau1}],
and that the main fingerprints of the phenomenon are
determined by the bulk properties of the material.

  It is interesting to note that the equilibrium screening
of $P$ by interfacial free charge is quite effective for any
thickness. 
The inset in Fig.~\ref{Ferro} assumes a value of $\chi_\eta=2$ for 
the plots of $\sigma$ and $P$ versus $d$, which was
chosen for illustrative purposes, but represents
a very small value for any real material. 
  Indeed, a more realistic value ($\chi_\eta=27$ for PbTiO$_3$)
 pushes the $\sigma$ curve right onto the $P$ curve, as
 shown in Fig.~\ref{Ferro}.

\begin{figure}
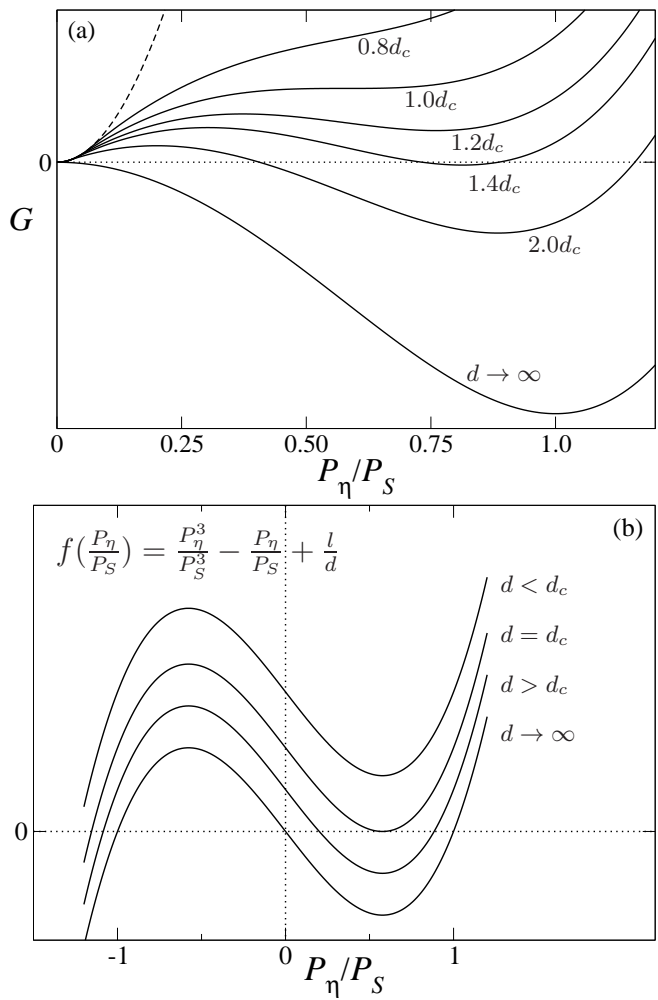

\begin{center}
\psfrag{0.8dc}{$0.8 d_c$}
\psfrag{1.0dc}{$1.0 d_c$} 
\psfrag{1.2dc}{$1.2 d_c$}
\psfrag{1.4dc}{$1.4 d_c$}
\psfrag{2.0dc}{$2.0 d_c$}
\psfrag{d_lim_infty}{$d\rightarrow\infty$}
\psfrag{d_larger}{$d>d_c$}
\psfrag{d_equal}{$d=d_c$}
\psfrag{d_smaller}{$d<d_c$}
\psfrag{equation}{\large$f({P_\eta \over P_S}) = {P_\eta^3 \over P_S^3}  - {P_\eta \over P_S}  + {l \over d}$}
\includegraphics[clip,width=\columnwidth]{GvsP_welec_limD_a}
\includegraphics[clip,width=\columnwidth]{GvsP_welec_limD_b}
\caption{\label{fig:GvsP} (a) Energy per unit volume as a function
of polarization for various thicknesses. The numbers next to the curves
indicate the values of the thickness in each case. Dashed line corresponds
to a solution with $\sigma=0$ while solid lines are the curves 
with $\sigma \neq 0$. Curves with $\sigma \neq 0$ possess an
equilibrium configuration only for $d \geq d_c$.
(b) Graphical solution of the equilibrium condition given by Eq. 
\ref{cubic2}.
}
\end{center}
\end{figure}
 
Fig. \ref{fig:GvsP} shows how the solution in Fig. \ref{Ferro} arises.
The energy functional of Eq. \ref{eq:Gfinal} has two possible
sets of solutions. For $\sigma=0$ the energy of the system as a 
function of the polarization is a parabola, as in a dielectric material,
and the equilibrium solution is $P_\eta = P =0$. 
%
Instead, the energy curve for $\sigma\neq 0$ [solid lines in
Fig. \ref{fig:GvsP}(a)] has extrema given by Eq. \ref{cubic2}. 
The function $f(p)=p^3-p$, corresponding to the limit of $d\rightarrow\infty$
of Eq. \ref{cubic2} and plotted in Fig. \ref{fig:GvsP}(b),
has roots at $-1$, 0, and 1. As $d$ is reduced, the cubic curve
shifts upward, then the upper root (corresponding to the polarization
of the equilibrium ferroelectric configuration) diminishes
while the middle one [corresponding to the energy bump in the 
$G(P_\eta)$ curves of Fig. \ref{fig:GvsP}(a)] becomes positive.
The consequence of the this is that an energy
barrier, which amplitude decays asymptotically as $d\rightarrow\infty$,
separates the paraelectric and ferroelectric configurations
for every $d>d_c$ and the system possesses three stable states: zero polarization
(insulating state) and the two opposite polarization orientations
(2D metallic state). 
Furthermore, as can be seen in Fig. \ref{fig:GvsP}(a), there is a range of thicknesses
$d_c<d<d_\mathrm{GS}\sim 1.4 d_c$ for which 
the ground state of the system is the paraelectric
configuration, and the ferroelectric one screened by the electronic reconstruction
is a local energy minimum. At the
critical thickness $d_c$ the minimum of the cubic curve 
touches the axes, meaning that at lower values of $d$ the
upper root becomes imaginary and ferroelectric configuration
ceases to be stable.

  The physical interpretation of the phenomenology described 
  above is clear: the appearance of 
spontaneous polarization in the film requires the screening 
of the depolarizing field. 
  This is accomplished by the accumulation of free charge 
 that results in the formation
of the corresponding 2DEG at the interface.
  Being the screening processes a surface effect and the tendency to
polarization a bulk effect, the latter dominates for thick enough 
films, while the former dominates in thin films. 
Interestingly the transition from paraelectric to ferroelectric,
or analogously, from an insulating to a conductive interface,
is discontinuous. This contrasts with the 
case of the LaAlO$_3$/SrTiO$_3$ interface for which an effective
model like this predicts a continuous transition with a gradual decrease of the
interface charge as the LaAlO$_3$ thickness is reduced.
\cite{Bristowe2014} This of course
cannot be observed experimentally since the thickness of LaAlO$_3$ can 
only be varied in units of the out-of-plane lattice constant, $c$. 
The continuous metal
insulator transition in SrTiO$_3$/LaAlO$_3$ 
can be shown using an external field to deplete charge from the 2DEG.
\cite{Thiel2006, Caviglia2008, Forg2012}
For the ferroelectricity-induced 2DEG 
we will confirm the discontinuous transition
under the application of an external electric
field in Sec. \ref{sec:efield}.

\subsubsection{Estimates}
   
   In addition to insights into the character of the 
transition, the model, still within the $g\rightarrow\infty$
approximation, 
allows estimations of the relevant 
magnitudes.
  As stated above the jump in polarization, $P_\eta^c$, is
$P_S/\sqrt{3}\sim 0.6 P_S$ and is thus quite universally 
defined, just dependent on the equilibrium polarization 
of the bulk ferroelectric material.

  For the critical film thickness, $d_c$, we can get
estimates by comparing the results obtained for the
LaAlO$_3$/SrTiO$_3$ interface \cite{Bristowe2014} and for the ferroelectric
film (Eqs.~\ref{lferro} and \ref{dcferro}),
\begin{equation}
d_c^{\rm LAO} = {(1+\chi^{\rm LAO})\over P_0^{\rm LAO} }
\varepsilon_0 \Delta^{\rm STO} \, ; \, \,
d_c^{\rm F} = {3\sqrt{3}\chi_\eta^{\rm F}\over P_S^{\rm F}} 
{\varepsilon_0 \Delta^{\rm F} } \, ,
\end{equation}
where F stands for ferroelectric, and $P_0^{\rm LAO}$ 
refers to half a quantum of polarization.
Assuming similar values of the band gap, and
considering now that the critical thickness for LaAlO$_3$ on
SrTiO$_3$ is around 4 perovskite layers, and that $3\sqrt{3}\sim 5$,
\begin{equation}
d_c^{\rm F} \sim 20 \left (\chi_\eta^{\rm F} /  \chi^{\rm LAO} \right ) \, \,
{\rm layers,}
\end{equation}
which can grow quite thick depending on how close
the temperature is to the bulk ferroelectric $T_c$.

\subsubsection{Effect of a finite density of states}
\label{sec:DOS}

\begin{figure}[]
\begin{center}
\includegraphics[clip,width=\columnwidth]{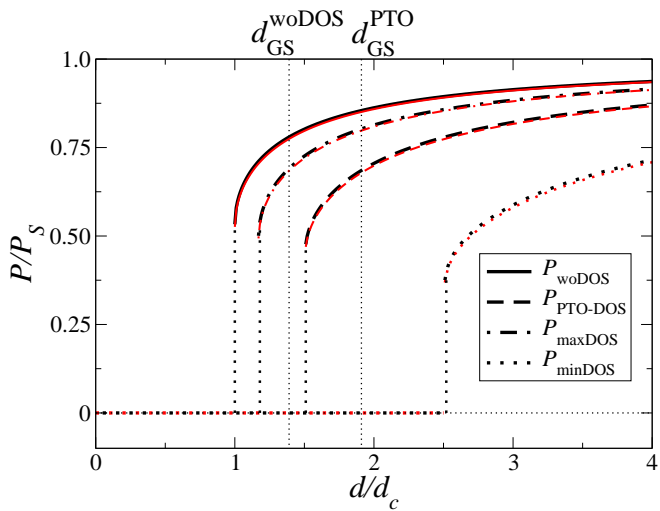}
\caption{\label{fig:Pvsd_wElecDOS} Total polarization
(black) and free charge (red) versus thickness 
calculated using different values for the DOS, $g$.
The different approximations for $g$ are $g\rightarrow\infty$ (solid lines), 
the value corresponding to
PbTiO$_3$ obtained from its bulk band structure \cite{Note1} (dashed), 
and the two extrema of an estimated range of values
for this family of materials (top limit in dashed-dotted and bottom
limit in dotted lines, respectively).
}
\end{center}
\end{figure}

For simplicity, we have assumed so far a large value of the 
reduced DOS, $g$. This approximation allowed us to neglect 
the energy cost of the filling of the 
valence and conduction bands, simplifying the analysis. 
The effect of a finite DOS
is to penalize the accumulation of free charge and consequently
it is expected to shift to larger thicknesses the metal-insulator
transition. 
To analyze how important the influence of a finite DOS is we plot 
in Fig. \ref{fig:Pvsd_wElecDOS} the curves of polarization as a function
of thickness for various values for the DOS, $g$.
For PbTiO$_3$ we estimated the corresponding 2-dimensional DOS for electrons and holes 
from the bulk band structure obtained from first principles. 
In addition to the usual considerations to calculate 
a DOS, in these systems one should also take into
account the fact that the interface lifts some degeneracies. We use the results 
of Ref. \onlinecite{Popovic2008} and \onlinecite{Delugas2011}, which showed
that after the electronic reconstruction the bottom of the conduction 
band in perovskite titanates has a $d_{xy}$ character, to estimate a DOS
for electrons of $g_e/e^2 = 1.2\cdot 10^{37}$ m$^{-2}$J$^{-1}$
($1.9\cdot 10^{14}$ cm$^{-2}$eV$^{-1}$ in more conventional units). We performed
a similar analysis for the valence band
\cite{Yin} 
to obtain the corresponding DOS for holes, 
which amounts to $g_h/e^2 = 2.5\cdot 10^{37}$ m$^{-2}$J$^{-1}$
($4.0\cdot 10^{14}$ cm$^{-2}$eV$^{-1}$).
Using the estimated DOS for PbTiO$_3$ we observe that there
is a significant increase of the transition 
thickness to about $\sim1.5 d_c$, where $d_c$ is given by 
Eq. \ref{dcferro} in the limit of $g\rightarrow\infty$.
The values of $d_c$ and
$d_\mathrm{GS}$ predicted by the model for PbTiO$_3$, both in the limit
of $g\rightarrow\infty$ and for a finite and realistic
DOS, can be found in Table \ref{tab:critThick}.
According to the values in Table \ref{tab:critThick}
the thickness necessary to produce a 2DEG in a ferroelectric thin film
is large compared with the case of the LaAlO$_3$/SrTiO$_3$ interface. 
For ferroelectrics grown under a high epitaxial strain, 
mechanisms for strain relaxation might start playing a role in this 
range of thicknesses, something that would affect the parameters in the model.
This is not the case for PbTiO$_3$ on SrTiO$_3$, the system
chosen for the numerical estimations throughout this 
article, for which epitaxy over thicknesses 
of several hundreds of unit cells can be achieved. \cite{Schlom2007}
Furthermore, the critical thicknesses in Table \ref{tab:critThick}
can be be potentially reduced by an appropriate choice of materials, and the 
model presented here provides a simple tool for the screening
of optimal systems.

\begin{table}[]
  \caption{Critical thickness for the stability of a 2DEG 
  induced by ferroelectricity, $d_c$, and for the polar phase
  to become the ground state with respect to a paraelectric 
  configuration, $d_{\rm GS}$. The thicknesses are calculated both in
  the limit of infinite DOS and using a realistic
  value for bulk PbTiO$_3$\cite{Note1}.
          }
  \begin{center}
    \begin{tabular*}{0.8\columnwidth}{@{\extracolsep{\fill}} lcc}
         \hline \hline
         &  $d_c$ (nm) & $d_{\rm GS}$ (nm) \\
         \hline
         without DOS & 5.5    & 7.7  \\
         with DOS    & 8.3   & 10.5 \\
         \hline
         \hline
      \end{tabular*}
   \end{center}
   \label{tab:critThick}
 \end{table}

Taking into account how large the effect of
the DOS can be, it is worthwhile to explore a range of 
sensible values, since different substrates, 
terminations or sources of free charge (see next Section) 
could give rise to very different values of $g$. 
If the source of free charge is, as assumed so far,
an electronic reconstruction and the materials involved in the 
formation of the interface are oxide perovskites, from the
estimated values $g_e$ and $g_h$ for PbTiO$_3$
we can consider that a given material of this family probably displays 
a DOS in the range between $5\cdot 10^{36}$ to $5\cdot 10^{37}$ m$^{-2}$J$^{-1}$
($8.0\cdot 10^{13}$ to $8.0\cdot 10^{14}$ cm$^{-2}$eV$^{-1}$).
These two values are used to calculate the two extra curves in
Fig. \ref{fig:Pvsd_wElecDOS}.
Inspecting Fig. \ref{fig:Pvsd_wElecDOS} one can see that
in fact materials with a large, but still reasonable, DOS
might show a dependence of the polarization with respect to
the thickness very close to the ideal situation. In materials
with a low DOS, on the other hand, the transition
may take place at thicknesses as large as $2.5d_c$ --- $3d_c$.

The strong influence of a finite DOS on the critical thickness 
for the onset of a 2DEG in ferroelectric films is specially
noteworthy because such dependence does not appear in
the case of the LaAlO$_3$/SrTiO$_3$ interface. In fact,
it was demonstrated in Ref. \onlinecite{Bristowe2014} that a model
for polar interfaces between dielectric materials 
analogous to the one presented here yields a critical thickness 
for the formation of the 2DEG that is independent of the DOS
(see Eq. 18 in the aforementioned article).

\subsubsection{Other sources of free charge}

Although the electronic reconstruction was the first mechanism for
the formation of the 2DEG in polar interface to be proposed, 
\cite{Nakagawa2006} alternative processes can also yield 
free carriers to screen the polarization discontinuity.
\cite{Bristowe2014}
Most notably, surface electrochemical processes such as 
adsorption of chemical species or formation of defects
induced by the internal electric field in the film 
are believed to play a major role in the screening 
at ferroelectric thin films
\cite{Fong2006, Spanier2006, Wang2009a, Stephenson2011, Bristowe2012a}
and polar interfaces.
\cite{Cen2008, Bristowe2011, Li2011a, Yu2014}
To account for this kind of mechanism Eq. \ref{eq:Ginitial} 
may be rewritten as
\begin{equation}\label{eqGredox}
G = U + {Cn\over d} + {\alpha n^2 \over 2d} 
 + {1 \over 2 \varepsilon_0} (nQ - P_\eta)^2 \; ,
\end{equation}
where $C$ is the formation energy of an isolated redox defect
in the absence of an electric field
and $Q$ is the charge provided by the defect. The defect-defect
interaction is accounted for in a mean-field approximation 
by the term 
$\alpha n^2/2d$.
It is easy to show that Eq. \ref{eqGredox}
and  Eq. \ref{eq:Ginitial} are in fact equivalent 
by simply making $\sigma = nQ$, $\Delta=C/Q$ and $\alpha=Q^2/g$. 
Indeed, the analysis 
presented above for electronic reconstruction
is parallel to any other planar charge screening
mechanism associated with an energy cost per 
surface/interface charge. 
This allows us to treat any equivalent screening mechanism
with the same equations, by considering the 
surface density of free charge, $\sigma$, and taking into account
the ``effective'' gap and DOS relevant for each process.
An analysis of the behavior under the simultaneous presence of more than one 
mechanism can also be done, as in Ref. \cite{Bristowe2014}
for LaAlO$_3$/SrTiO$_3$

\subsubsection{Assuming given $\sigma$ or $P$}

  The analysis in the previous subsections assumes equilibrium, 
and thus neglects any kinetic effects, which can be very important,
e.g. in the formation of defects or the Zener tunneling of the
carriers across the film.
  In some occasions such kinetic effects may dominate.
  We can easily consider the situation in which a certain 
concentration of redox defects $n$ has been generated on
the surface, e.g. at growth, which are then frozen in.
Such scenario corresponds to a ferroelectric layer in open
boundary conditions with fixed electric displacement $D$, where
$D=\sigma=nQ$. In this situation $\sigma$ is not a variable 
but the parameter determining the electrostatic boundary
conditions of the system. The relevant free energy is now
simply \cite{Stengel2009c}
\begin{equation}
 G = U+{1 \over 2\varepsilon_0}(\sigma-P)^2.
\end{equation}
Using again the transformations for $P$ given by 
Eq. \ref{eq:Pe} and \ref{eq:P} we get
\begin{equation}
 G = {1 \over 2\varepsilon_0 \chi_\eta}
     \left({1\over 4} {P_\eta^4 \over P_S^2} - {1 \over 2}P_\eta^2\right) 
     + {1 \over 2\varepsilon_0 \varepsilon_\infty} \left(\sigma - P_\eta\right)^2.
 \label{eq:Gsigma}
\end{equation}
  We can then ask what would be the equilibrium polarization
$P$ for given values of $\sigma$.
  Minimizing $G$ with respect to $P$ gives the equation
\begin{equation}
{P_\eta^3 \over P_S^3} + \left({2 \chi_\eta \over \varepsilon_\infty} -1\right) 
{P_\eta\over P_S}
- {2\chi_\eta \sigma \over \varepsilon_\infty P_S}=0 \, ,
\end{equation}
where the last term is a constant. For any 
$\chi_\eta/ \varepsilon_\infty>1/2$ this
equation has a single real root, which is positive.
 Considering for simplicity a large value of $\chi_\eta/\varepsilon_\infty$, and for
values of $\sigma$ not much larger than $P_S$, the
solution can be approximated by
\begin{equation}
P_\eta \sim \sigma + {\varepsilon_\infty\over 2\chi_\eta} 
\sigma\left [ 1 - \left ({\sigma \over P_S} \right )^2 \right ] \, .
\end{equation}
which using Eq. \ref{eq:P} transforms into
\begin{equation}
P \sim \sigma + {1 \over 2\chi_\eta} 
\sigma\left [ 1 - \left ({\sigma \over P_S} \right )^2 \right ] \, .
\label{eq:fixedSigma}
\end{equation}

 That is, the polarization responds by compensating the
effective polarization given by the 
2DEG carriers (first term) except for a small deviation, which
is positive ($P>\sigma$) for $\sigma<P_S$ and negative for $\sigma >P_S$,
or, in other words, the polarization tends to screen the field 
generated by the fixed surface/interface charge, 
but with a slight tendency towards $P_S$.
  Note that under fixed $D$ boundary conditions 
the free energy of the system scales with the volume 
(it does not have surface terms) and thus
the behavior obtained is independent of
film thickness.
Several works \cite{Wang2009a, Bristowe2012a}
have demonstrated that the manipulation the surface chemistry 
can be used to switch the polarization of a ferroelectric.
Furthermore, the first principles simulations presented in 
Ref.~\onlinecite{Bristowe2012a} showed how the polarization 
of a BaTiO$_3$ film followed the charge density set by 
charged defects at the surface,  as predicted by Eq. \ref{eq:fixedSigma}.

  Similarly one could ask what would be the equilibrium
concentration of free charge if the polarization $P$ had
been frozen in by some mechanism.
  In such (unlikely) case, we would minimize Eq.~\ref{eq:Gfinal}
with respect to $\sigma$ for fixed $P_\eta$ and then use 
Eq. \ref{eq:P}, obtaining
\begin{equation}
\sigma= P - {\varepsilon_0 \Delta \over d}
\end{equation}
which gives a phenomenology very similar to the cases
of the LaAlO$_3$/SrTiO$_3$ interface, in which the
polarization is also fixed (in that case to half a quantum)
\cite{Bristowe2014}.

\subsection{First principles simulations}

In order to test the validity of the model, its predictions can be compared with 
results obtained from first principles simulations of ferroelectric thin
films.
Bearing in mind that first principles simulations are typically 
performed at zero temperature, the results obtained with this method
should be compared with the low temperature limit of the model. 
Nevertheless, as long as the temperature is relatively far from the transition one,
the phenomenology should be qualitatively the same.

The study presented in Ref. \onlinecite{Fredrickson2015}, showing that 
electronic reconstruction can stabilize a spontaneous polarization in 
symmetrical BaTiO$_3$/SrTiO$_3$ heterostructures, 
constitutes the first argument supporting the model.
In addition to this, here we perform additional
DFT calculations on a model system consisting of a 
slab of PbTiO$_3$ (PbO terminated on both sides) in vacuum,
An in-plane lattice constant of 3.874 {\AA} was chosen to mimic the strain
of a SrTiO$_3$ substrate (which is not explicitly included in the calculation).
Even though such geometry is not representative of typical 
experimental devices, and properties of the 2DEG such as its confinement 
or the mobility of the charge carriers would be very different
from more realistic samples like those depicted in Fig. 
\ref{fig:geom}(c) and \ref{fig:geom}(d), we choose here
this simple test-case with the sole purpose of illustrating
some of the basic predictions of the model. 
%
%
%
The calculations were performed within the local density 
approximation, using the {\sc siesta} code.\cite{Soler2002}  Reciprocal 
space integrations were carried out on a Monkhorst-Pack
\cite{Monkhorst1976, Moreno1992}
$k$-point grid equivalent to $6\times6\times6$ in a 
five-atoms perovskite unit cell. For real space integrations 
a uniform grid with an equivalent plane-wave cutoff of 600 Ry 
was used. A dipole correction was introduced to avoid spurious
interaction between periodic images of the slab along the out of plane
direction.
Initial coordinates were generated stacking $m$ unit cells 
of ``bulk-strained'' PbTiO$_3$ in the ferroelectric phase. 
Starting from the polar configuration we expect that during relaxation 
the system will remain in the metastable configuration predicted by the model
for $d>d_c$.
Then, all the atomic coordinates of the slabs were relaxed until the forces 
were smaller than 0.04 eV/\AA. 

In Table \ref{tab:firstPrinciples} we list the equilibrium 
polarization and energies of the resulting structures as a function
of the thickness. It was found that for all slabs with 
$d\leq12$ unit cells the atoms moved back to the
centrosymmetric positions and the system was insulating. 
Instead, for $d\geq14$ unit cells the slab remained polar
and the surfaces were metallic.
Within the range of thicknesses
analyzed here, the energy of the system in the polar phase is 
higher than in the paraelecric one, confirming that the 
ferroelectric/2DEG configuration is still metastable, but the energy
difference decreases rapidly with increasing thickness suggesting that
$d_{\rm GS}$ should be around 10 nm.

\begin{table}[]
  \caption{Evolution of the polarization and the energy with respect to the 
  non polar phase for PbTiO$_3$ slabs in vacuum. Here we report the values of the 
  polarization of stable structures after geometry optimizations
  initialized in a polar configuration, for $d\leq 12$ the system 
  spontaneously goes back to the paraelectric phase during the relaxation.
  Energies are given per formula unit.
          }
  \begin{center}
    \begin{tabular*}{\columnwidth}{@{\extracolsep{\fill} } cccc}
         \hline \hline
         $d$ (unit cells) &  $d$ (nm) & $P/P_S$ 
         & $G_{\rm FE} - G_{\rm para}$ (meV) \\
         \hline
         10 & 4.0 & 0 & -  \\
         12 & 4.8 & 0 & -  \\
         14 & 5.6 & 0.52 & 43 \\
         16 & 6.4 & 0.62 & 32 \\
         18 & 7.2 & 0.79 & 25 \\
         \hline
         \hline
         \end{tabular*}
   \end{center}
   \label{tab:firstPrinciples}
 \end{table}

We can now obtain the 
predicted critical thickness for this material according to the model. 
The four parameters, $\chi_\eta$, $P_S$, $\Delta$ and $c$, needed 
to estimate the critical thickness were independently 
computed from DFT calculations on ``bulk-strained'' PbTiO$_3$
and found to be 27, 0.78 C/m$^2$, 1.6 V and 4.03 {\AA}
respectively. 
The fact that the model parameters were obtained from first principles 
calculations means that they also represent the low temperature limit
and a direct comparison with the first principles simulations of the 
slab can be made.
Using Eqs. \ref{dcferro} and \ref{lferro}
one gets $d_c =2.4$ nm $\sim 6$ unit cells, a value significantly smaller
than the one estimated from the first principles
simulations. Nevertheless, if a reasonable value of the 
DOS is used ($g_e/e^2 = 1.2 \cdot 10^{37}$ and 
$g_h/e^2 =2.5 \cdot 10^{37}$ m$^{-2}$J$^{-1}$ for electrons and holes, 
as in Fig. \ref{fig:Pvsd_wElecDOS}) the critical
thickness increases up to 5.1 nm $\sim 13$ unit cells, in excellent
agreement with the simulations.

\section{Switching the 2DEG with an external applied field}
\label{sec:efield}

The main interest of having a 2DEG in a ferroelectric thin film is that
the polar discontinuity at the interface might be manipulated 
in non trivial ways with the application of an external electric field. 
Since the formation of a 2DEG relies on the presence of a 
polarization mismatch at the interface, the standard geometry for
ferroelectric capacitors, with the bottom electrode deposited between the
substrate and the ferroelectric film cannot be used. 
Instead, in order to preserve
the polarity of the interface, the electrodes for the 
manipulation of the 2DEG should be placed
underneath the dielectric substrate and on top of the ferroelectric
surface (the latter can be the tip of an atomic force microscope), adopting
the field-effect transistor geometry often used to electrically tune the 
2DEG at the LaAlO$_3$/SrTiO$_3$ interface 
\cite{Caviglia2008,Forg2012,Hosoda2013} (see for instace Fig. 1 of
Ref. \onlinecite{Forg2012}).

One obvious possible application for this system is the non volatile 
switching of the 2DEG at the interface. 
If the two interfaces or
surfaces of the ferroelectric thin film are equivalent, switching the 
direction of the polarization
would simply exchange the 2DEG and 2DHG between opposite interfaces.
If instead, the interfaces
are dissimilar (if one of them is actually a surface, for instance) 
switching the polarization
might also change the effective band gap $\Delta$, modifying the value of the critical
thickness $d_c$. If the thickness of the ferroelectric layer is close to $d_c$, 
switching the polarization would then induce a 
metal-insulator transition at the interface.
This however requires the application of large electric fields to 
be able to switch the polarization of the ferroelectric layer, even larger than
for the bulk material since the free charge of the 2DEG might respond to the 
electric field and screen it. Nevertheless, as we will see here, switching the
polarization is not the only way to turn on and off the 2DEG at a
ferroelectric interface.

To evaluate the effect of an external electric field in 
a ferroelectric thin film with a 2DEG at one of its interfaces
we extend the model introduced in previous sections,
adding to Eq. \ref{eq:Gfinal} a new term corresponding to the 
interaction of the uncompensated
polarization with the external field,
\begin{equation}
 G = U + {\sigma\Delta\over d} + \frac{\sigma^2}{2gd} + 
    \frac{1}{2\varepsilon_0}(\sigma-P)^2 
                + (\sigma-P)\mathcal{E}.
 \label{eq:energy_E}
\end{equation}
In this expression a positive value of $\mathcal{E}$ represents 
an electric field parallel to and with the same sign of the polarization.
It should be noted that $\mathcal{E}$ is an \emph{external} electric field
and the total field experienced by the ferroelectric is
$\mathcal{E}_\mathrm{FE}=\mathcal{E} + (\sigma -P)/\varepsilon_0$, 
with the correct sign criterion. 
The choice of the electric field as the independent variable is in this case
natural, since the dependence of the polarization on $\mathcal{E}$
only involves the characteristics of the ferroelectric layer. 
In experiments, however, the variable that can be directly controlled
is typically a gate voltage. In such case, the dependence of the 
polarization with the gate voltage requires also knowing 
details about the substrate. 
Nevertheless, given the specific
details of a device, the relation between $\mathcal{E}$
and a gate voltage can be obtained through
\begin{equation}
 V = \mathcal{E}\left({d_\mathrm{I}\over \varepsilon_\mathrm{I}} + d\right)
     - {P-\sigma \over \varepsilon_0}d,
 \label{eq:voltage}
\end{equation}
where $d_\mathrm{I}$ and $\varepsilon_\mathrm{I}$ are the thickness and relative
permittivity of the insulating substrate. With this expression one can estimate,
for instance, that for a 300 nm thick SrTiO$_3$ substrate,
with a relative permittivity of 300, the maximum electric field considered in this
section ($0.4 P_S/\varepsilon_0$) requires the application of 
approximately 30 V between the top and bottom electrodes.
 
\begin{figure}[]
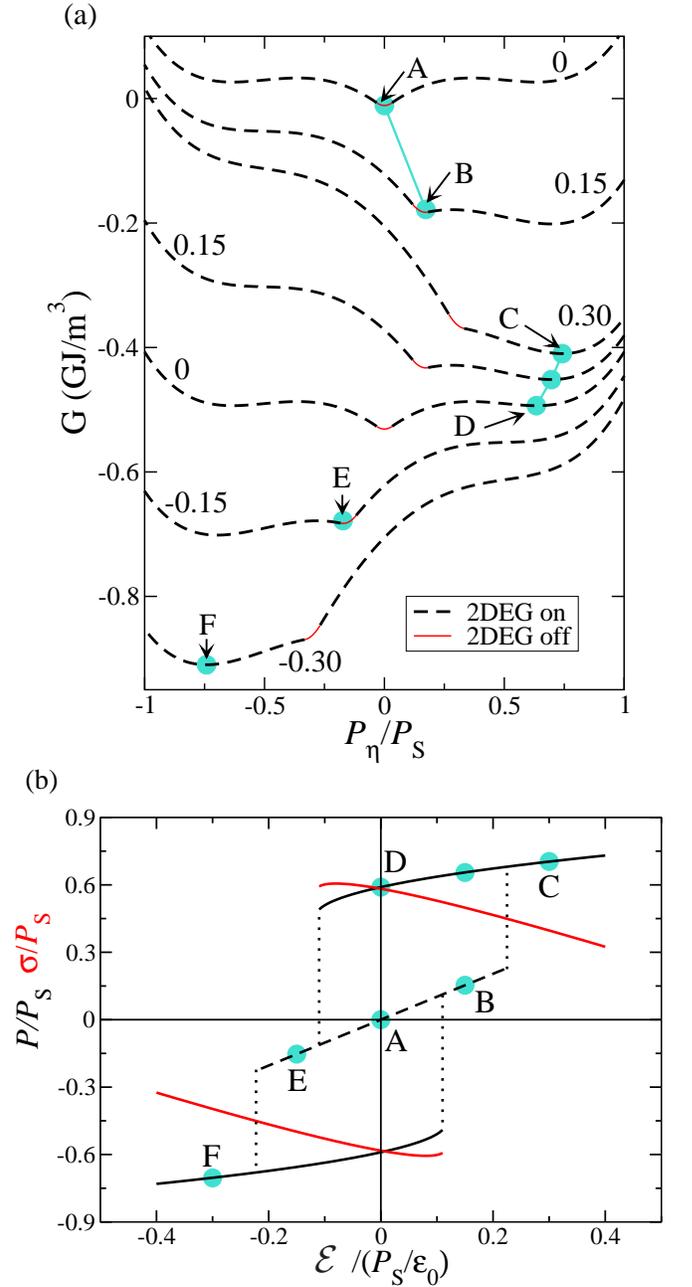

  \begin{center}
    \psfrag{EF}{\Large $\mathcal{E}$}
    \includegraphics[clip, width=.9\columnwidth]{GvsP_evolWithE}
    \includegraphics[clip, width=\columnwidth]{hysteresis_9nm}
  \end{center}
  \caption{(a) Energy curves for a ferroelectric thin film as a function
  of the polarization for different values of the applied electric field
  (numerical labels next to the curves, in units of $P_S/\varepsilon_0$). 
  Red solid (black dashed) sections correspond to solutions with $\sigma=0$
  ($\sigma\neq 0$). Energy curves have been shifted vertically for clarity.
  Light blue dots indicate equilibrium states as $\mathcal{E}$
  is swept. Connected (disconnected) dots represent a continuous change 
  (jump) in $P_\eta$.
  (b) Hysteresis loop for $P$ (black line) and $\sigma$
  (red line) as a function of the applied electric field.
  For the polarization, solid lines represent 
  the ferroelectric state with the 2DEG while the
  dashed lines correspond to the paraelectric phase. 
  Both (a) and (b) are calculated for a PbTiO$_3$ thin film 
  with a thickness of 9 nm ($d\gtrsim d_c\sim8.3$ nm). 
  Points labeled in the top panel correspond to those indicated
  in bottom panel, please find a detailed description in the text.
            }
  \label{fig:externalField}
\end{figure}

The equilibrium polarization of the ferroelectric under an applied
electric field $\mathcal{E}$ is found after writing Eq. \ref{eq:energy_E}
in terms of $P_\eta$, using again Eq. \ref{eq:P}, 
and imposing the equilibrium condition 
$\partial G/\partial P_\eta = \partial G/\partial \sigma = 0$.
To analyze the evolution of the polarization and free
charge as a function of the 
applied electric field, it is important to recall that 
the energy curves plotted in Fig. \ref{fig:GvsP} are actually
the result of merging two curves corresponding to two different 
sets of solutions, one for solutions with $\sigma=0$
and another one for solutions with $\sigma\neq0$. 
In Fig. \ref{fig:externalField}(a) we plot as red solid lines
and black dashed lines 
the curves corresponding to $\sigma=0$ and $\sigma\neq0$,
respectively, for a 10 nm thick
PbTiO$_3$ film and various values of the 
applied electric field.
Since Eq. \ref{eq:energy_E}
is only valid for $\sigma>0$, the sections of the curves corresponding to
an electric field antiparallel to the screening field due to $\sigma$ 
($\sigma>0$, $\mathcal{E}<0$ or $\sigma<0$, $\mathcal{E}>0$) 
are all calculated with $\sigma>0$ and $\mathcal{E}<0$
and reversed with respect to $P_\eta=0$ when necessary.

As discussed before, the energy curve at zero field has the appearance of
a symmetric \emph{triple-well} profile, with two metastable
ferroelectric states (with 2DEG) and a central paraelectric
minimum [see curve at the top in Fig. \ref{fig:externalField}(a)].
As with the typical ferroelectric double-well energy landscape,
the effect of the external field is to tilt the energy curves,
modifying the relative stability of the different equilibrium
configurations and the potential barriers. 
The energy curves in Fig. \ref{fig:externalField}(a)
demonstrates that the application of an external field
can be used to switch between the paraelectric
(without 2DEG) and polarized (with 2DEG) states. 
Most interestingly, the tri-stability of the energy curves
in Fig. \ref{fig:externalField}(a) suggests that the metal-insulator
transitions should display a rather complex hysteresis. In the following
lines we use the energy curves in Fig. \ref{fig:externalField}(a) to understand
the shape of the hysteresis loop depicted in Fig. \ref{fig:externalField}(b),
obtained using the parameters that correspond to a 9 nm thick
PbTiO$_3$ thin film. 
As shown in Fig. \ref{fig:externalField}(a), 
at zero electric 
field both the paraelectric ($\sigma=0$, $P_\eta=0$) and
ferroelectric ($\sigma \sim P_\eta \sim 0.6P_S$) configurations are stable, 
with the former being the ground state of the ferroelectric film
[point A in panels (a) and (b) of 
Fig. \ref{fig:externalField}]. 
As we increase the magnitude
of the external field, the minimum of the $\sigma\neq 0$ curve deepens,
eventually becoming the most stable phase of the system. However there
is a potential barrier separating the two stable phases, thus for small fields
($\mathcal{E}\lesssim0.2 P_S/\varepsilon_0$), starting in a 
configuration with $\sigma=0$, the system can remain in the paraelectric
phase (B). Nevertheless, for a high enough field the system
eventually overcomes the potential barrier and the 
monodomain configuration as well as the 2DEG are switched on (C). 
If then the
electric field is decreased a potential barrier prevents 
the transition back to the paraelectric phase (D). The switching 
takes place for negative fields for which the polar 
configuration is no longer stable (E).
For this particular thickness (9 nm), when the energy curve for 
$P>0$, $\sigma\neq0$ loses its minimum,
there still exist an energy barrier separating the $\sigma=0$ state
from the one with $P<0$ and $\sigma\neq0$,
thus the system remains in the insulating phase (E).
Therefore, for this thickness, the switching of the polarization and
surface free charge polarity occurs through the 
non-polar phase.
Eventually, for large enough fields the system switches again to 
a ferroelectric state with a metallic interface (F). 

Assuming the typical geometry depicted in Fig. \ref{fig:geom}(c), an ideal
\emph{interface} free of defects, and the 
widely accepted situation in which free charge at the \emph{surface}
gets trapped by defects or adsorbed molecules, the reversal of the polarization
implies a switching between a 2DEG and a 2DHG at the buried interface. 
This is an interesting result, because a 2DHG has never been observed in 
LaAlO$_3$/SrTiO$_3$. The absence of conductivity at the $p$-type interface
is commonly attributed to the fact 
that the polarity of the AlO$_3$/SrO boundary is screened by
defects formed during the growth process. For the system 
discussed here, after the deposition the interface
is buried and protected from further redox reactions.
If the ferroelectric is polarized down as grown,
one would expect to initially find the 2DEG, 
which is less susceptible to
be screened by defects, at the interface, 
then a switching of the polarization might 
be able to induce the formation of the elusive 2DHG.
Since electrons and holes can and usually do present very different 
characteristics (such as mobility), this possibility 
constitutes an interesting opportunity for the 
design of new electronic devices based on oxide interfaces,
where one could not only play with the on and off switching of the
2DEG but also with the switching between different gases.

\begin{figure}[]
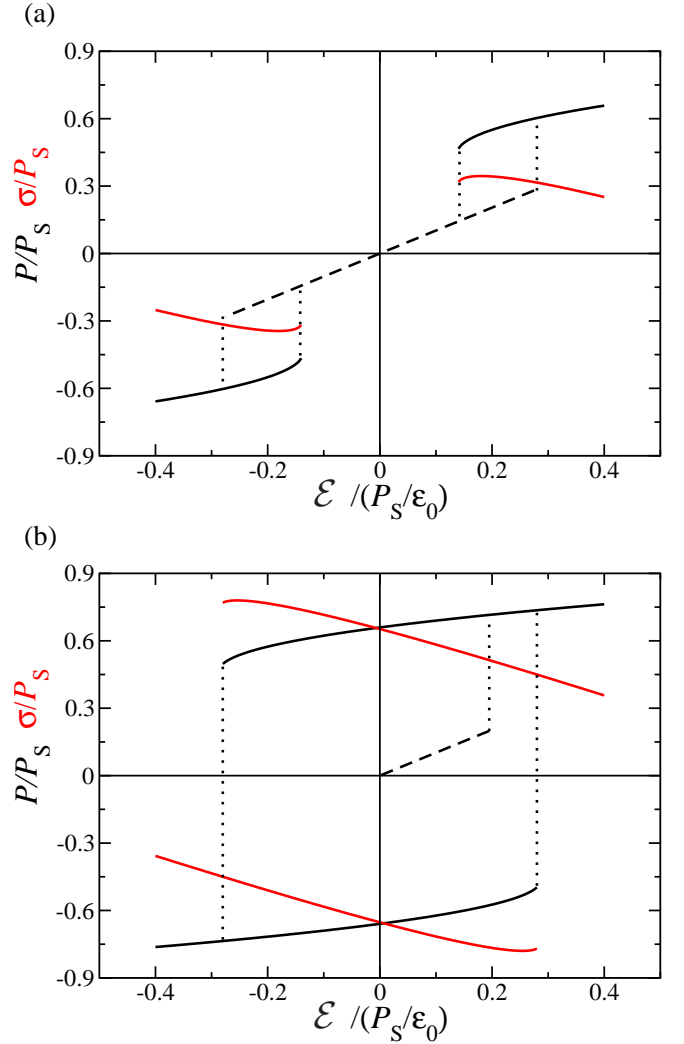

  \begin{center}
    \psfrag{EF}{\Large $\mathcal{E}$}
    \includegraphics[clip, width=\columnwidth]{hysteresis_7p5nm}
    \includegraphics[clip, width=\columnwidth]{hysteresis_10nm}
  \end{center}
  \caption{Hysteresis loop for the polarization (black line) and surface free charge
  (red line) as a function of the applied electric field 
  for a ferroelectric thin film of two different thicknesses:
  (a) 7.5 nm ($d<d_c\sim 8.3$ nm), and (b) 10 nm 
  (representative of the $d\gg d_c$ situation).}
  \label{fig:hysteresis}
\end{figure}

\begin{figure}[]
  \begin{center}
    \includegraphics[clip,width=\columnwidth]{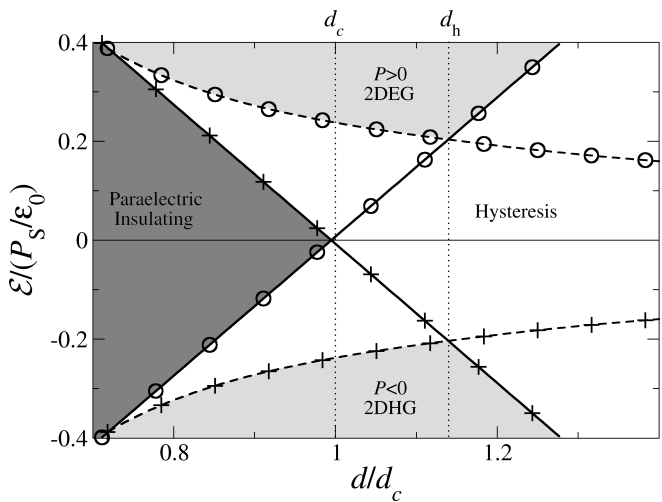}
  \end{center}
  \caption{Phase diagram as a function of thickness $d$ and external 
  electric field $\mathcal{E}$ for a PbTiO$_3$ thin film. 
  Circles (crosses) indicate an upward (downward) jump in polarization 
  in forward (backward) sweep. Solid (dashed) lines correspond to
  transitions from $\sigma\neq0$ ($\sigma=0$) to $\sigma=0$ ($\sigma\neq0$). 
  Regions of paraelectricity and ferroelectricity (with 2DEG or 2DHG)
  coexistence) are shown in dark and light gray respectively. In white, 
  different states are accessible depending on the sweeping history. 
  }
  \label{fig:phasediag}
\end{figure}

Another interesting aspect of this system is the fact
that the shape of the hysteresis
loop is strongly dependent on the thickness of the ferroelectric film.
Fig. \ref{fig:hysteresis}(a) illustrates the case of a ferroelectric
film with a thickness below the critical one for the stability of the 
polar configuration at zero field. In this case the switching displays
two separate loops centered at $|\mathcal{E}|>0$. The electric field
can be used to induce the transition from insulating to metallic interface, 
but the 2DEG would be volatile.
This hysteresis loop resembles the one corresponding to an
antiferroelectric, displaying a phase transition to a polar state induced by an
external electric field and an absence of remnant polarization
at zero field. Such features maximize the electrostatic energy that 
can be stored in a capacitor and suggests that hysteresis loops in
ferroelectric thin films might be tailored and optimized  
for energy storage applications. \cite{Ortega2012}
In the opposite limit
of large thicknesses, shown in Fig. \ref{fig:hysteresis}(b),
once the $\sigma\neq0$ state has been reached 
the switching takes place directly between polar states, i.e. 
between a 2DEG and a 2DHG.

The whole phase diagram for PbTiO$_3$ thin films as a function of 
$d$ and $\mathcal{E}$, including the hysteresis regions, is shown in
Fig. \ref{fig:phasediag}. There, the regions of paraelectricity, and 
ferroelectricity (with 2DEG or 2DHG coexistence) are shown in dark and 
light gray, respectively. In white, different states are possible depending
on the sweeping history. For $d<d_c$, at zero field, the ferroelectric
state is not stable and two separate hysteresis loops are observed at finite field, 
as in Fig. \ref{fig:hysteresis}(a). In Fig. \ref{fig:phasediag}, 
$d_h$ marks the thickness above which the switching occurs directly 
between the two polar states, like in Fig. \ref{fig:hysteresis}(b). In
the region between $d_c$ and $d_h$ all three situations, ($\sigma\neq0$, $P>0$),
($\sigma\neq0$, $P<0$), and ($\sigma=0$), are accessible, like in 
Fig. \ref{fig:externalField}. The phenomenology presented in 
Fig. \ref{fig:phasediag} should be quite general, 
but the shapes of the different
boundaries between regions of the phase diagram
depend on the parameters of the material, most notably
on the DOS, $g$. 

At this point it is worth noting that the effect of a finite DOS 
in the model cannot be neglected to get the right dependence of 
polarization and free charge with the electric field. In the limit
of $g\rightarrow\infty$, starting from the paraelectric phase 
and as the electric field increases, the system would eventually
switch to the ferroelectric/2DEG state. However, once in the
ferroelectric/2DEG state, if the electric field is reversed,
since there is no penalty for $\sigma$ to grow indefinitely 
(its dependence with the electric field is linear, as in the $\sigma\Delta$
term), the system would never switch back to paraelectric or to 
the opposite polarization state. In fact, under the
$g\rightarrow\infty$ approximation the shape of energy curve
of the $\sigma\neq0$ configurations does not change with the application
of an electric field (it only shifts up or down), therefore 
the energy minimum corresponding to the
ferroelectric state is a metastable configuration
for any value of the electric field.

We have assumed here that the free charge can freely move from
one interface or surface of the ferroelectric to the other, 
implying that no potential barriers are involved in
these charge transfer processes. 
This might be reasonable for very thin ferroelectric films
(where tunneling between the two interfaces is easy) and 
if the potential barriers for redox reactions at
the surface are relatively low.
Instead, if after the formation
of the 2DEG the screening charge cannot respond to an external electric field --
this can be the case of free charge created after an electronic
reconstruction in a relatively thick film --
the polarization would be pinned by the free charge
and the ferroelectric would behave as a linear dielectric with a very small
susceptibility.
This problem may be overcome by contacting the interface with electrodes, 
puncturing through the ferroelectric layer. The metallic contacts    
could act as the source of free charge for the modulation of the 
2DEG at the interface.

\section{Discussion}
\label{sec:Discussion}
\subsection{Competition with polarization domains}
\label{sec:poly}

In the analysis presented so far we have assumed that no 
polarization domains are formed within the ferroelectric.
This competing screening mechanism, however, constitutes the main 
obstacle for the formation of a 2DEG in a ferroelectric thin film.
Both the electronic reconstruction or the redox processes
are possible sources of screening that can help 
to stabilize monodomain phases in ferroelectric thin films where alternative
mechanisms (such as metallic electrodes) are not present. 
Nevertheless, the breaking into 
polarization domains competes
with the processes discussed here, since
in a fully compensated polydomain configuration
the net polarization 
charge at surfaces or interfaces is zero,
eliminating the driving force for an eventual 
electronic reconstruction or surface electrochemical processes.

Experiments on ferroelectric thin films grown on insulating substrates
are abundant in the literature. In the following lines we review
the most relevant experimental literature in order to 
find potential test cases where the hypothesis presented
here could be investigated.
 We only discuss 
here those experiments where the ferroelectric material is grown directly in
contact with the insulator, since the presence of a 
buffer metallic electrode would screen the polar discontinuity at the
interface.
\subsubsection{PbTiO$_3$/SrTiO$_3$}

The most widely studied ferroelectric heterointerface is probably 
the case of PbTiO$_3$ thin films on SrTiO$_3$ substrates.
In Refs. \onlinecite{Thompson1997} and \onlinecite{Bedzyk2000}, for instance,
authors used x-ray diffraction methods to determine the polarization
distribution of PbTiO$_3$ films on SrTiO$_3$. In Ref. \onlinecite{Thompson1997}
a 10 nm film was found to be monodomain as grown with the polarization pointing 
down (i.e. towards the substrate). 
Instead, in Ref. \onlinecite{Bedzyk2000} a transition with thickness 
was observed from polydomain (10 and 20 nm thick films) to monodomain 
with the polarization pointing up 
(40 nm thick films).
Refs. \onlinecite{Streiffer2002, Fong2004a} investigate the phase
diagram of PbTiO$_3$ thin films with respect to temperature
and thickness.
Ferroelectric films with thicknesses
ranging from 0.4 (1 unit cell) to 42 nm were grown on SrTiO$_3$ substrates and the 
polarization configuration was explored using x-ray scattering. 
At room temperature polydomain phases were observed for 
thicknesses as small as 1.2 nm (3 unit cells).
For $d\gtrsim 2$ nm satellite peaks in the x-ray scattering
maps disappear but tetragonality is consistent with polar
PbTiO$_3$, suggesting a transition to a monodomain phase. 
Atomic force microscopy measurements were also consistent with a transition from
polydomain to monodomain in thicker films. \cite{Thompson2008}

Using coherent Bragg rod analysis D. D. Fong \textit{et al.} were able
to obtain a real space mapping of the atomic positions
(and thus of the polarization) in PbTiO$_3$/SrTiO$_3$
interfaces. \cite{Fong2005} The authors analyze PbTiO$_3$ films with thicknesses
of 1.6 and 3.6 nm that, depending on the cooling process after growth, can be
stabilized either in a mono (slow cooling down to room temperature)
or a polydomain phase (fast cooling to 181 $^\circ$C).
The technique used in this work is 
sensitive to the local electronic density and a priori could
be used to image the free charge accumulated
to screen the polar catastrophe at the interfaces. 
However this would require comparing
the electronic density of a single sample in the monodomain and
polydomain configuration (for which no electronic transfer is expected). 
Unfortunately in the paper poly and monodomain phases could only be compared 
\emph{in two different samples} and any difference in the surface termination
and interface intermixing obscures the possible presence of screening charge.

Despite the discrepancies about the transition thickness for poly to
monodomain phase (that might be ascribed to different growth and 
characterization conditions)
the phase diagram that is obtained from this collection of experiments
is consistent with the one provided in 
Refs. \onlinecite{Fong2004a,Thompson2008}.
This phase diagram results from the compilation of all the experiments 
discussed above and shows that, at room temperature, for very small thicknesses 
films are paraelectric and as the thickness is increased
the ferroelectric films evolve from paraelectric to
polydomain and then (in a ``sluggish transition'', in the words of the authors)
to a monodomain phase.
This suggests that films are polydomain
immediately after deposition but became monodomain as they are cooled down
(unless quenching is used to freeze the  polydomain structure). 
The driving force for the transition is possibly the 
fact that the orientation of the polarization 
affects the reaction energies of the relevant redox 
processes at the surface, what might make one set of domains
more energetically favorable than the other.

\subsubsection{BaTiO$_3$/SrTiO$_3$}

In Ref. \onlinecite{Tenne2009} authors study BaTiO$_3$ films grown on 
SrTiO$_3$ substrates. SrTiO$_3$ exerts a compressive strain on BaTiO$_3$
stabilizing a tetragonal phase. In this work, a combination of UV Raman 
spectroscopy and synchrotron x-ray scattering is used to
test the polarity of films with thicknesses of 1.6 to 10 nm. 
It is found that films of all thicknesses are polar at low temperature. At 
room temperature, films capped with a 10 nm SrTiO$_3$ layer were polydomain
with regular domain sizes. X-ray scattering spectra of uncapped films
did not show satellite peaks but authors were not able to 
establish whether this was due to nonperiodic domains, domain sizes larger
than the experimental resolution or stabilization of a monodomain phase.
The model proposed here might offer a possible interpretation of this experiment. 
The uncapped samples could sustain a monodomain phase screened by the 
accumulation of free charge at the interface and surface. In the capped 
films there would be two possible ways for stabilizing a monodomain configuration. 
(i) Redox processes at the SrTiO$_3$ surface could provide the necessary 
free charge to screen the polar discontinuity at the ferroelectric/substrate interface, 
but then the capping SrTiO$_3$ layer should be polarized as well, with the 
corresponding energy cost. (ii) Electronic 
reconstruction within the ferroelectric layer could simultaneously screen the 
polar discontinuity at both interfaces (with the electrode and the capping layer), 
but the value of $\Delta$ for this process
is much larger than for the surface electrochemical reactions. In either case, 
the energy cost of a monodomain polarization 
in the capped film would be higher than for the uncapped one, 
consistent with the observation of polydomain phases 
in the capped samples. 

In Ref. \onlinecite{Dubourdieu2013a} authors report ferroelectricity in
BaTiO$_3$/SrTiO$_3$/SiO$_2$/Si heterostructures. Even without a top electrode, authors
claim that polarization can be written in BaTiO$_3$ films as thin as 8 nm
with the tip of an atomic force microscope. For thicknesses
of 1.6 nm or below, written domains were unstable. They also report
hysteresis loops in the piezoelectric response without top electrode.

\bigskip
\noindent 
All these experiments, and specially the phase diagram for PbTiO$_3$ films on
SrTiO$_3$ substrates provided in Ref. \onlinecite{Thompson2008},
consistently support the possibility of stabilizing a 
monodomain phase in ferroelectric thin films grown directly 
on insulating substrates.
Unfortunately the source of screening required to stabilize such
configuration is discussed in very little detail and the possibility
of an electronic reconstruction or alternative process that could form a 
2DEG or 2DHG at the interface was not contemplated in those works. 
Very recently, however, the formation of a 2DEG at the interface between CaZrO$_3$
and SrTiO$_3$ has been reported. \cite{Chen2015} Even if CaZrO$_3$ is not
polar in bulk, it is argued in Ref. \onlinecite{Chen2015} that the 
compressive strain exerted by the substrate might induce a polarization
of the CaZrO$_3$ which would be responsible for the formation of the 2DEG. 
Although ferroelectricity (switchable polarization) could not be demonstrated,
the evolution of the free charge with thickness 
is in good agreement with the model presented here.  
Further characterization of electromechanical and transport properties  
in samples similar to those discussed in this section,
and the comparison with the results of the model
would be critical to assess the hypothesis presented here. 

\subsubsection{A model for the competition with polarization domains}

The model discussed in Sec. \ref{sec:Formation} takes into account the
balance of the monodomain ferroelectricity with a paraelectric
phase, but cannot say anything about the competition with a polydomain
phase. Since experiments demonstrate that such a competition
exists and the balance is delicate, here we use a simple
model to investigate the relative stability of these two phases.

Since the two screening mechanisms are mutually exclusive 
we do not need to add new terms to our energy expression of Eq. \ref{eq:Gfinal}.
Instead, to analyze the competition between the two phases (monodomain and polydomain)
we need to compare the thickness evolution of the energy of a thin
film in the two different scenarios. 
For the polydomain phase we assume a $180^\circ$ domain structure, in which 
straight stripes of the material, all with the same width in the direction perpendicular
to the domain wall,
have an out of plane polarization of the same magnitude but with 
alternating orientations.
For such idealized version of the 
domain structures typically found in tetragonal ferroelectric thin films
the energy per unit of volume of the polydomain phase can be expressed as 

\begin{equation}
 G_{\rm poly} = U + {\Sigma \over w} + G_{\rm elec},
 \label{eq:energy_poly1}
\end{equation}

\noindent where $\Sigma$ is the energy per unit of area of a domain wall and $w$ 
is the domain width.
The electrostatic energy $G_{\rm elec}$ due to 
stray fields in the polydomain configuration
is proportional to the domain width, \cite{Kittel1946, Mitsui1953}
$G_{\rm elec} = {\gamma w / d}$, where the proportionality constant can be calculated 
to be 

\begin{equation}
 \gamma = \frac{8.416 P^2}{\pi^3\varepsilon_0
           [1+(\varepsilon_x\varepsilon_z)^{1/2}]},
\end{equation}

\noindent for $180^\circ$ stripe domains. \cite{Ozaki1995} The width of domain walls
in a typical ferroelectric is vanishingly small and remnant electric
fields in polydomain configurations decay exponentially away from the
surface and the domain wall, thus, 
except in the limit of thicknesses of a few unit cells,
we can approximate $|P|\sim P_S$ throughout the film and $U = U_0(P_S)$ as a
constant. Using this result and differentiating 
Eq. \ref{eq:energy_poly1} with respect to $w$ to find the 
equilibrium domain width for a given thickness of the film, one obtains the well
known Kittel law \cite{Kittel1946}

\begin{equation}
 w^2 = \frac{\Sigma d}{\gamma}.
 \label{eq:Kittel}
\end{equation}

\noindent Substituting this expression for the equilibrium domain width into 
Eq. \ref{eq:energy_poly1} leads to the following
expression for the energy per unit volume

\begin{equation}
 G_{\rm poly}(d) = U_0(P_S) + 2\left( {\Sigma \gamma \over d} \right)^{1/2}.
  \label{eq:energy_poly2}
\end{equation}

\noindent In Eq. \ref{eq:energy_poly2} the first term is negative and 
constant, independent of the thickness of the film; the second
one is positive, diverges at $d\rightarrow 0$ (due to the divergence
of the domain wall density given by the Kittel law) and decays with
the thickness of the ferroelectric.

\begin{figure}
  \begin{center}
    \includegraphics[clip,width=\columnwidth]{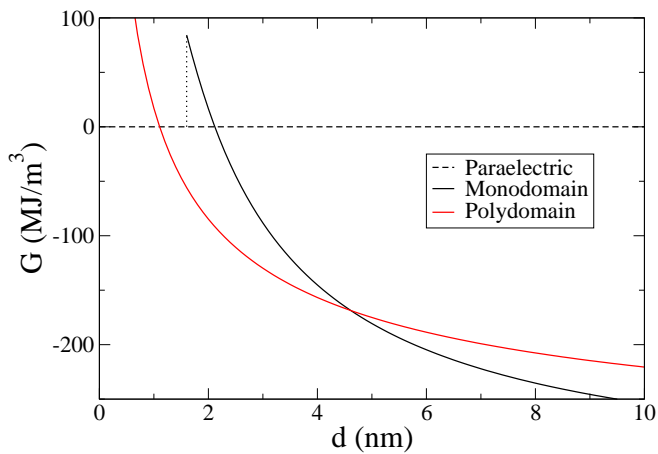}
  \end{center}
  \caption{Energy density of a ferroelectric thin film in monodomain 
           (black curve) and polydomain (red curve) states. Curves
           are calculated for bulk PbTiO$_3$, in the limit of $g\rightarrow\infty$,
           using parameters obtained from 
           first principles calculations. \cite{Note1} In this plot $\Delta = 1$ V,
           realistic for surface redox processes.
            }
  \label{fig:polyVSmono_energy}
\end{figure}

The Kittel law has been shown to be valid down to thicknesses of
a few nanometers in typical ferroelectric thin films. 
\cite{Catalan2007, Catalan2008}
This simplified expression for the energy
is expected to break down below the limit of a few unit cells,
at which point one should take into account the 
finite width of the domain wall, the stray fields
and the inhomogeneities of the polarization inside the domains.
Neglecting these effects, and using again the parameters obtained 
from first principles simulations for PbTiO$_3$, \cite{Note1}
we compare the energy of the polydomain configuration 
of Eq. \ref{eq:energy_poly2} (red curve in Fig. \ref{fig:polyVSmono_energy})
with that of a monodomain 
phase where we allow the possibility of 
having surface charge, as discussed in Sec. \ref{sec:theModel}, 
given by Eq. \ref{eq:Gfinal} (black curve in Fig. \ref{fig:polyVSmono_energy}).
As demonstrated before, ferroelectricity in a monodomain phase
becomes stable only above a critical thickness, $d>d_c$. In 
Fig. \ref{fig:polyVSmono_energy} we are plotting the case where 
$\Delta = C/Q \sim 1$ V, corresponding to a screening by surface 
redox processes, that yields a value of $d_c=1.60$ nm. 
Interestingly, even though for small thicknesses the model 
predicts that the most favorable scenario is the formation 
of polarization domains, it also shows that there should be
a crossover between the two phases, such that above a critical
thickness the stabilization of a monodomain state by the 
formation of a 2DEG would become energetically favorable 
over the breaking into domains of polarization.
For the chosen parameters this crossover takes 
place for a thickness of about 4.6 nm, well within the
range of thicknesses that are typically grown in ferroelectric
thin film experiments.
The thickness at which the 
transition from a polydomain to a monodomain configuration takes place
strongly depends on the value of $\Delta$. This parameter takes different 
values depending on the screening mechanism we are considering: it is
simply the band gap of the ferroelectric if we are assuming an electronic
reconstruction scenario in a free standing slab, and $\Delta=C/Q$ 
if we consider the possibility of having electrochemical
processes taking place at the free surfaces of the material. 
In the first case it is obvious that the band gap underestimation 
by the traditional exchange-correlation
functionals would greatly affect the estimated crossover thickness obtained
from DFT calculations, as shown in Table \ref{tab:PTO}. In the case of a 
ferroelectric thin film on top of a substrate the relevant gap $\Delta$
would depend also on the band gap of the substrate material and the
band alignment at the interface, as indicated in Fig. \ref{fig:geom}.

\begin{table}[t]
  \caption{ Estimated thickness for the mono to polydomain 
            crossover, $d_{\rm c-o}$, in PbTiO$_3$ thin films. 
            Parameters for the model
            have been obtained from first principles simulations. \cite{Note1}
          }
  \begin{center}
    \begin{tabular*}{0.8\columnwidth}{@{\extracolsep{\fill}} lcc}
         \hline \hline
         &  $\Delta$ (V) & $d_{\rm c-o}$ (nm) \\
         \hline
         Redox processes & $C/Q\sim 1$  & 4.6  \\
         LDA gap         & 1.45         & 10 \\
         Exp. gap        & 3.6          & 64 \\
         \hline
         \hline
      \end{tabular*}
   \end{center}
   \label{tab:PTO}
 \end{table}

PbTiO$_3$ was chosen as an example to test 
the model presented here because it is a
prototypical ferroelectric and has good characteristics to exhibit the 
formation of the 2DEG at reasonable thicknesses, as demonstrated by the 
values in Tables \ref{tab:critThick} and \ref{tab:PTO}.
But in fact one could now use Eq. \ref{eq:energy_poly2} and \ref{eq:Gfinal} 
to explore scenarios that could favor even further the 
formation of a 2DEG over the domains of polarization.
Ideally one would like to find a material with large domain
wall energy and spontaneous polarization and small band gap
and dielectric constant. A good candidate could be the so called 
``super tetragonal'' phase of BiFeO$_3$ 
(at least from a theoretical point of view, experimentally 
the stabilization of these phases requires a large in-plane
compressive strain and the material usually forms a mixed phase
with other monoclinic phases of BiFeO$_3$; \cite{Zeches-09}
another disadvantage of this material is that experimental samples
often present a very large leakage, behaving like a semiconductor more than like a
true insulator). First principles simulations predict that for
in-plane compressive strains larger than $\sim 5\%$ 
some phases of BiFeO$_3$ could display a spontaneous 
polarization \cite{Hatt2010a, Dieguez2011} of up to 150 C/m$^2$ 
and domain wall energies \cite{Ren2013} of about 250 mJ/m$^2$.
A band gap of 3.1 eV has been obtained by optical absorption 
measurements in these highly strained phases. \cite{Chen2010}
All these parameters would yield an estimated crossover
thickness for the transition from poly to monodomain of
8.7 nm in the electronic reconstruction scenario. Furthermore, 
if the monodomain polarization is stabilized by 
surface electrochemical reactions this phase is
more favorable than the polydomain for any thickness,
as shown in Table \ref{tab:BFO}.

\begin{table}[t]
  \caption{ Estimated thickness for the mono to polydomain
            crossover in thin films of BiFeO$_3$ under a compressive
            strain of $5\%$. The parameters for the model
            have been obtained from first principles simulations available in the
            literature (see text) except for the relative permittivity,
            that was calculated here ($\varepsilon_x=35$, $\varepsilon_z=25$). 
            Formation of a 2DEG by means of surface redox processes is 
            more favorable than a polydomain configurations for all thicknesses.
          }
  \begin{center}
    \begin{tabular*}{0.8\columnwidth}{@{\extracolsep{\fill}} lcc}
         \hline \hline
         &  $\Delta$ (V) & $d_{\rm c-o}$ (nm) \\
         \hline
         Redox processes & $C/Q\sim 1$ & 0\\
         Exp. gap        & 3.1         & 8.7 \\
         \hline
         \hline
      \end{tabular*}
   \end{center}
   \label{tab:BFO}
 \end{table}
 
This result suggests that even if the breaking up into domains
is a very effective mechanism for the screening of the polar discontinuity
at the surface or interface in a ferroelectric thin film, the formation of
a 2DEG is indeed viable and might form for an appropriate 
combination of materials and boundary conditions. 
The prediction, using such a simple model,
of a transition from polydomain to monodomain as the 
thickness is increased is in good agreement with the experimental
observations and should constitute a further motivation to explore
the scenario proposed here.
 
\subsection{Ferroelectric substrate}

\begin{figure}
  \begin{center}
    \includegraphics[clip,width=\columnwidth]{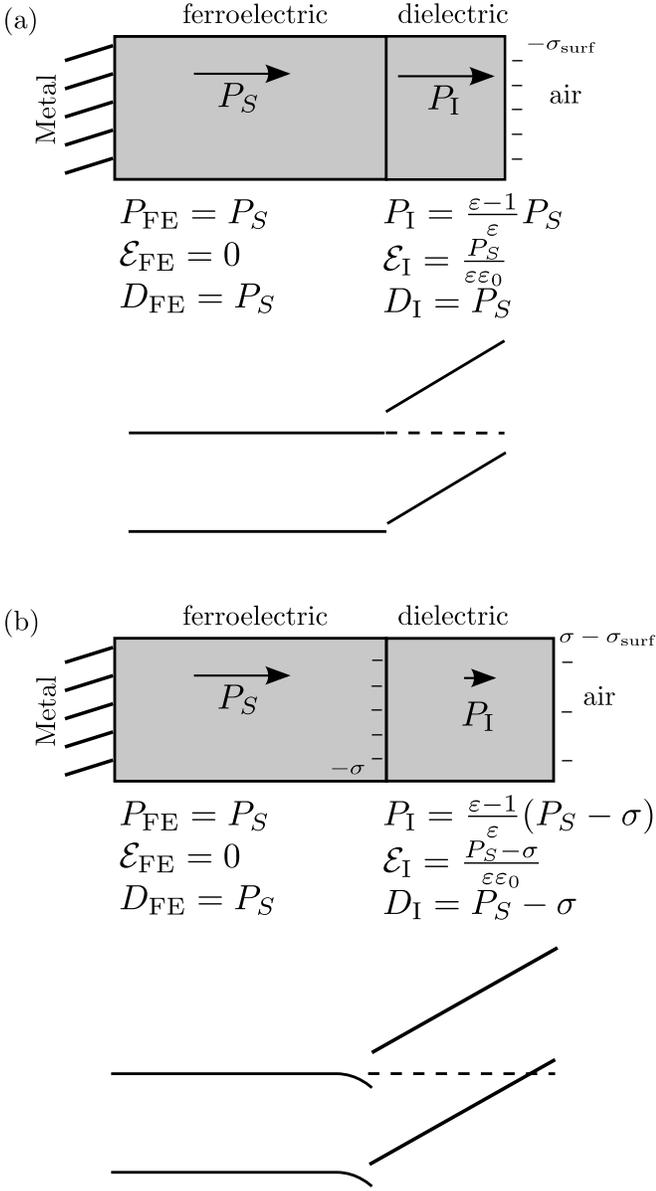}
  \end{center}
  \caption{Schematic depiction of the electrostatics involved in the case
  of an insulating thin film on top of a ferroelectric substrate, both (a)
  before and (b) after an electronic reconstruction. 
  A schematic band diagram of the system is shown at the bottom of each panel.
  Note that for this geometry, even before the electronic reconstruction, free charges 
  need to be present at the surface to screen the polar discontinuity 
  with the vacuum/air region.
            }
  \label{fig:FEsubstrate}
\end{figure}

  If the ferroelectric material is the substrate, and the film
is a dielectric perovskite the situation is different.
  If the substrate is thick and is 
connected to a metal at the back, the polarization discontinuity 
at the ferroelectric/metal interface would be screened and the 
ferroelectric would display a strong tendency to develop a finite
polarization. 
This, however, gives rise to a polarization mismatch 
at the ferroelectric/dielectric over-layer interface
that needs to be compensated to fully screen the depolarizing field.
In fact, the problem shares many similarities with the
LaAlO$_3$/SrTiO$_3$. In the latter case, the thick substrate 
imposes a $D=0$ electrostatic boundary conditions. Since
LaAlO$_3$ has a finite polarization at zero electric field, the $D=0$ 
condition implies that an electric field develops inside the polar
material. For a thin LaAlO$_3$ film, the material polarizes under the action
of the field, tending to reduce the polar discontinuity. Above a 
critical thickness though, an electronic reconstruction (or a more
complex mechanism possibly involving redox processes) becomes more
energetically favorable and a 2DEG forms at the interface. 
Instead, in the case of a non polar dielectric layer on top 
of a thick ferroelectric substrate, the thick ferroelectric
imposes $D=P_S$. As in the case of the LaAlO$_3$, in the absence of free charge
at the interface this condition
induces an electric field in the insulating top layer, which for small
thicknesses will polarize [see Fig. \ref{fig:FEsubstrate}(a)] until, 
as the thickness increases, 
the energy balance favors formation of a 2DEG at the interface
[see Fig. \ref{fig:FEsubstrate}(b)]. 
There are two important differences with respect to the LaAlO$_3$/SrTiO$_3$ case though.
In the case of the ferroelectric substrate/dielectric film
a polar discontinuity exists also at the free surface of the
dielectric film, which requires the accumulation of some 
superficial free charge (most likely provided by chemical adsorbates,
schematically represented by negative signs in Fig. \ref{fig:FEsubstrate},
in accordance with the choice of polarization orientation in the ferroelectric) 
even before the screening at the interface sets in. 
But most importantly, in this system $\Delta P$ 
can be changed (switched) by an electric
field and by temperature.

%

   Consider for example an SrTiO$_3$ film on a BaTiO$_3$ substrate.
In this case $\Delta P$ is the bulk polarization of the substrate,
(at low $T$ analogous in magnitude to the polarization mismatch in
LaAlO$_3$/SrTiO$_3$),
and there would be then an instability with a similar
critical thickness of the film for either electronic reconstruction,
the appearance of redox defects, or both. 
  Assuming that beyond that critical thickness equilibrium
is established in the presence of a bulk polarization of BaTiO$_3$
parallel to $z$, a 2DEG should appear
at the interface, of electrons for one sign of BaTiO$_3$'s 
polarization, of holes for the other. 
  If switching the substrate ferroelectric, and assuming 
equilibration, the gas of electrons should transform into 
gas of holes and vice-versa.
  Alternatively one could think of a 2DEG (or 2DHG) being
switched on and off with $T$ if, say, starting from BaTiO$_3$
above the ferroelectric critical $T$, and letting it cool down
until it polarizes enough to give rise to the gas.

  Finally, if one is interested in switching on and off the
2DEG with a transversal electric field, a polar film can
be considered, such as LaAlO$_3$, and a ferroelectric substrate
chosen such that its bulk polarization at the temperature
of operation is close to the LaAlO$_3$ half-quantum. 
  With both polarizations aligned, $\Delta P \sim 0$
and no 2DEG should arise. 
  If the ferroelectric is then switched, $\Delta P$ becomes
approximately a whole quantum, and the 2DEG should 
be strongly populated and stable for quite a thin film,
possibly allowing for the Zener tunneling to take place
populating the 2DEG.
  In this geometry, for either configuration 
there would still be a polarization
discontinuity at the free surface, but this would probably 
be screened by redox processes.

\subsection{Hyperferroelectrics}
\label{sec:otherFE}

\begin{figure}[]
  \begin{center}
    \includegraphics[clip,width=\columnwidth]{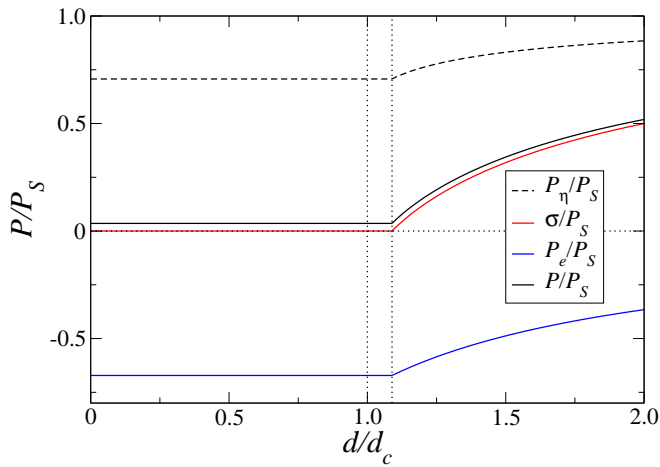}
  \end{center}
  \caption{Total polarization (black solid line), zero field contribution
           (dashed), electronic contribution (solid blue)
           and surface free charge (solid red)
           as a function of thickness. $d_c$ is the one
           given by Eq. \ref{dcferro} and in this case it does not
           mark the critical thickness for 2DEG formation.
            }
  \label{fig:hyper}
\end{figure}

Recently a new family of ferroelectric materials has been discovered 
that are predicted to display a finite polarization at $D=\sigma=0$, 
hence the reason why they were named \emph{hyperferroelectrics}.
\cite{Garrity2014}
Their capacity to display a finite polarization even in the absence of
any source of screening might give the impression that the tendency 
of this materials to display 2DEG is even stronger than for traditional
ferroelectrics. For this reason, in this Section we present an
analysis for these materials similar to the one carried out 
in Sec. \ref{sec:theModel}.
Hyperferroelectrics are
fundamentally equivalent to conventional ferroelectrics in the sense that the 
ferroelectric phase transition is driven by an unstable polar mode. Even though 
their behavior in open circuit boundary conditions is radically different from that
of conventional ferroelectrics, the underlying physics is completely analogous. 
The fundamental differences are the low effective charges associated with the polar
mode and the large polarizability of the electronic cloud. The consequence of 
this is that in open boundary conditions the electronic polarization screens the 
zero-field contribution and a large fraction of the polar distortion remains stable.
As demonstrated in Ref. \onlinecite{Garrity2014} 
these materials are expected to display a small but finite remnant polarization at
$D=0$.

Since the basic mechanism for ferroelectricity in hyperferroelectrics is the same as 
for traditional ferroelectrics, the expression of the free energy for 
these materials is again Eq. \ref{eq:Gfinal}. One can in fact, use that expression
to find the condition for a ferroelectric to behave as a hyperferroelectric
by calculating the equilibrium polarization for $\sigma=0$. 
Using Eq. \ref{eq:Gsigma} with $\sigma=0$ and differentiating with respect
to the polarization, one obtains the equilibrium condition
\begin{equation}
 {P_\eta^3 \over P_S^3} + {P_\eta \over P_S}
 \left({2\chi_\eta\over\varepsilon_\infty} - 1\right) = 0
\end{equation}
This equation has solutions
\begin{equation}
 \label{eq:condHyper}
 P_\eta^{D=0} = 0; \qquad
 P_\eta^{D=0} = \pm\sqrt{1 - {2\chi_\eta \over \varepsilon_\infty}}.
\end{equation}
The non-zero solutions are real only if $\varepsilon_\infty>2\chi_\eta$,  
which constitutes the condition for hyperferroelectricity.

Here we perform an analysis completely analogous to the one in
Sec. \ref{sec:theModel} to obtain 
the evolution of polarization and free charge in a thin film of a 
hyperferroelectric.
We use the set of parameters corresponding to 
LiBeSb reported in Ref. \onlinecite{Garrity2014} ($\chi_\eta\sim6$ is estimated
from the curvature at the minimum of the double well potential).
The result is plotted in Fig. \ref{fig:hyper}.
Three important differences are observed in the curves of Fig. \ref{fig:hyper}
with respect to the corresponding ones for a traditional ferroelectric
in Fig. \ref{Ferro}. First, as expected for these materials, 
for a thickness below
the onset for the electronic reconstruction the material displays
a large polar distortion that is effectively screened by the
electrons, resulting in a small but finite total polarization
at $D=\sigma=0$. Second, the 2DEG becomes stable at a thickness larger than
the critical one given by Eq. \ref{dcferro}. Finally, neither
the total polarization nor the free charge display a 
discontinuous jump at the transition. In fact the second and third 
observations are connected. Using Eq. \ref{eq:dG/dsigma},
\ref{dcferro} and \ref{eq:Pc} one can demonstrate that the value 
of the free charge at $d_c$ is 
\begin{equation}
 \sigma_c = {P_S \over \sqrt{3}}
 \left(1-{\varepsilon_\infty \over 3\chi_\eta}\right).
\end{equation}
This expression yields negative values of $\sigma$ for 
$\varepsilon_\infty>3\chi_\eta$, which are not a valid solution of
the model. Therefore for those materials with $\varepsilon_\infty>3\chi_\eta$, 
like LiBeSb, $\sigma$ goes to zero (the 2DEG vanishes)
in a continuous transition and 
for a thickness larger than $d_c$. 
The range of $2\chi_\eta<\varepsilon_\infty<3\chi_\eta$
constitutes a third regime in the phase diagram with respect to
$\varepsilon_\infty/\chi_\eta$, where
the material is a hyperferroelectric but still displays a discontinuous
jump in the polarization and the free charge.

As demonstrated in Ref. \onlinecite{Garrity2014},
the small but finite polarization at
$D=\sigma=0$ is a consequence of the strong 
screening provided by the large electronic polarizability, 
but the ferroelectric instability is not
necessarily stronger than in normal ferroelectrics.
We have seen in this section that the use of
hyerferroelectrics does not favor the formation of a 2DEG
as compared with a traditional ferroelectric, and in fact some 
of the most remarkable features of this system might be lost, such as 
the discontinuous transition of the polarization. 

\section{Conclusions} 

We have used a simple model to demonstrate that under
the appropriate conditions a monodomain out of plane polarization
may be stabilized in a ferroelectric thin film grown
directly on an insulating substrate
through the formation of a 2DEG at its interface. 
Although there are important analogies with 
the related polar interfaces between dielectric materials,
of which LaAlO$_3$/SrTiO$_3$ is the prototypical example,
there are striking differences in behavior too.
For the 2DEG at ferroelectric interfaces 
the model predicts that a discontinuous transition 
as a function of thickness takes
place between the paraelectric (without 2DEG) and the ferroelectric
(with 2DEG) phases, with an abrupt jump in both polarization and 
free charge. 
Also in contrast with the LaAlO$_3$/SrTiO$_3$, we have demonstrated
that the thickness for this transition strongly depends 
on the DOS of the 2DEG.

One of the key features that was sought in this system was the ability
to switch on and off the 2DEG as well as between 
a 2DEG and a 2DHG with the application of an external electric field.
The model shows a complex hysteresis behavior, an effect that poses
interesting possibilities for energy storage and non-volatile
memory applications.

The model has also been used to discuss possible strategies
to favor this state over the competing paraelectric and
polydomain configurations. 
We hope that the predictions derived from the model,
supported by the first principles simulations and 
some of which agree with many features 
from available experiments, will motivate the search for 
2DEG in these systems.

\section{Acknowledgments}
We acknowledge computing 
resources of CAMGRID in Cambridge, DIPC in San Sebasti\'an and the 
Spanish Supercomputer Network (RES).
  This work has been partly funded by 
  MINECO-Spain (Grant FIS2012-37549-C05),
  UK's EPSRC,
  and the ARC project TheMoTherm (Grant No. 10/15-03).
  Work at Argonne supported by DOE-DES under Contract No.
  DE-AC02-06CH11357.
  PhG acknowledges a Research Professorship of the 
  Francqui Foundation (Belgium), and NCB a research
  fellowship from the Royal Commission for the 
  Exhibition of 1851 and support from the Thomas Young Centre under grant TYC-101.

%

\end{document}